\documentclass[journal]{IEEEtran}
\usepackage{ifpdf}

\usepackage{cite}

\ifCLASSINFOpdf
  \usepackage[pdftex]{graphicx}
  \graphicspath{{figs/}{}}
  \DeclareGraphicsExtensions{.pdf,.jpeg,.png}
  \usepackage{epstopdf}
\else
  
  \usepackage[dvips]{graphicx}
  \graphicspath{{figs/}}
  \DeclareGraphicsExtensions{.eps}
\fi
\usepackage[cmex10]{amsmath}
\usepackage{array}

\usepackage[normalem]{ulem}	

\usepackage{verbatim}

\usepackage{color}

\usepackage{version}

\excludeversion{notes}

\begin{document}
%
\title{Calibrated Single-Contact Voltage Sensor for High-Voltage Monitoring Applications}
%
%
%


\author{Jake~S.~Bobowski, 
        Saimoom~Ferdous, and~Thomas~Johnson,~\IEEEmembership{Member,~IEEE}
\thanks{Manuscript submitted on \today.}
\thanks{J.S.~Bobowski is with the Department
of Physics, University of British Columbia, Kelowna,
BC, V1V 1V7 Canada (e-mail: \mbox{Jake.Bobowski@ubc.ca}).}
\thanks{T. Johnson and S. Ferdous are with the School of Engineering, University of British Columbia, Kelowna, BC, V1V 1V7, Canada (e-mail: \mbox{Thomas.Johnson@ubc.ca}).}
}

%
%


\markboth{Bobowski \MakeLowercase{\textit{et al.}}: Calibrated Single-Contact Voltage Sensor for High-Voltage Monitoring Applications}{}
%



\maketitle

\begin{abstract}
A single-contact voltage sensor designed for accurate measurements of ac voltages across a pair of conductors is described. The sensor design is motivated by remote monitoring applications where accurate voltage measurement of high-voltage transmission lines is required.  The body of the sensor is electrically and mechanically attached to a single conductor: either the neutral or high-voltage conductor. A capacitive sensing plate attached to the sensor creates a capacitive voltage divider using the stray capacitance to the non-contacted line. A very high-impedance buffer is used to measure the voltage across the divider output and estimate the line voltage. An important part of the work includes a method of calibrating the sensor such that blind voltage measurements can be made without knowing the exact geometry of the conductors. Other important aspects of the design include a two-stage voltage divider for retaining accuracy and increasing the voltage range of the sensor. The work is supported by extensive numerical simulation models which were used to determine the optimum design for the sensing plate and to evaluate the sensitivity to different configurations including conductor spacing and the height above ground. For calibration values which are accurate to 1\%, the line voltage can be measured with an accuracy of 10\%.  The paper describes the theory, design, and experimental verification of the sensor up to a line voltage of 7.5~kVrms.  
\end{abstract}


%
\IEEEpeerreviewmaketitle

\section{Introduction}\label{sec:intro}
%
%
%
%

\IEEEPARstart{A}{s} smart grid technology evolves, utility companies continue to improve the monitoring capabilities in the grid to optimize the delivery of power to network loads. Sensors to measure voltage and power factor provide real-time feedback to a central monitoring station where dynamic changes to voltage and reactive power compensators can be made to reduce loss. 
Wireless monitoring of end user loads using smart meter technology is now common, and substation monitoring has always been available. However, real-time monitoring on distribution lines between substations and loads is not common. The distribution grid is extensive and distribution line monitoring requires a cost-effect sensor network strategy. Desirable features of a sensor unit  include easy installation and a wireless link to back haul the remote measurement data to a central office.

The research work in this paper is motivated by the challenge of making accurate voltage measurements at intermediate points in the distribution grid. Power factor measurements  require accurate phase measurements between voltage and current; however, the measurement can be made without knowing the exact magnitude of the voltage. Since voltage control \cite{2011_Redfern_Univ_Power_Eng_Conf_smart_grid_voltage_control,
2013_Willoughby_IEEE_Smart_Grid_Newsletter} in the grid is becoming increasingly important to reduce energy loss in the distribution of electricity, accurate measurements of the potential difference between conductors is required.

A key constraint in the work is to design a sensor that is easily deployable. The voltage sensor is designed to be integrated into a measurement unit that is clipped to a single wire with no contact to any other wires. Current measurements are easily made using a Rogowski coil that can be integrated into a clamp which attaches the sensor unit to the wire. On the other hand, an accurate voltage measurement from contact with a single wire is much more difficult.  In this work, {our goal was} to design a voltage sensor capable of accurately determining the absolute potential difference between a pair of conductors while contacting only one of the conductors~\cite{2014_Johnson_Voltage_Sonsor_Patent}.

A review of the literature does not yield too many papers which investigate accurate single-contact voltage measurements for high-voltage transmission applications. In a Master's thesis by van der Merwe \cite{2006_vanderMerwe_MSc_thesis_three_phase_cap_voltage_sensor}, he presents a study and experiments for {measuring the} line potential in a three-phase system using capacitive coupling concepts.  In his work, calibration of unknown capacitances is found from either  numerical simulations for a specific wire geometry  or  experimentally by applying known potential differences between conductors to  determine the value of the capacitance matrix. In our work, we seek to find \emph{in situ} calibration methods where unknown capacitances must be estimated without exact knowledge of the wire geometry or without the need to apply test potentials between the wires. The constraints arise from the practical deployment of voltage sensing technology in a smart grid where installation must occur directly on live systems with arbitrary wire configurations.

Another three-phase voltage sensor configuration is described in Ref.~\cite{1998_Ohchi_IEE_Gen_Trans_Dist_coaxial_voltage_sensor_analysis}. The authors call the sensor a static induction type of voltage sensor and it consists of two plates near the conductor. In one design, the plates are arranged as a coaxial capacitor structure completely enclosing the conductor and in another design a semicircular conductor and a flat plate are used. The paper focuses on the evaluation of the phase errors detected by three sensors and the results assume the geometry and dielectric structure between the sensor and conductor are known.

In other work \cite{2011_Chan_Sensors_Actuators_A_Phys_noncontact_capacitive_sensor}, a non-contact voltage sensor employing two capacitive sensing plates is described. 
The sensor is tested on an isolated conductor and requires a known reference voltage to be applied to the system to calibrate and find coefficients for a fitting function. A series of measurements with different displacements from the conductor are also made to estimate the potential of the conductor. In the design which we report on, neither calibration voltages nor changes in displacement are used for calibration.

Other work related to the measurement of potential using electric field sensors includes biomedical applications \cite{2007_Cauwenberghs_IEEE_BIOCAS_conf_ECG_EEG_noncontact} and inspection of active current-carrying conductors in integrated circuits \cite{2011_Lerch_Sensors_Actuators_A_Phys_contactless_inspection_circuits}. In these applications, the electric fields are very small and  a significant challenge is amplifying and detecting small changes in potential difference. {These methods focus on detecting relative changes in potential whereas our work focuses on methods to measure absolute potential differences without direct contact to both conductors.}



{In our design the principle concept is to construct a calibrated capacitive divider from which accurate measurements of voltage can be made. The divider consists of a controlled and known capacitance between the contacted line and a sensing plate and a second unknown stray capacitance from the sensing plate to the non-contacted line.  Challenges in this design include developing calibration methods to estimate the unknown stray capacitance to the non-contacted line and optimizing the physical design of the sensor to maximize sensitivity.} An experimental prototype has been constructed and evaluated in a high-voltage test bed. The results are encouraging and{, assuming a calibration accuracy of 1\%,} absolute voltage measurements with an accuracy of $\pm$~10\% were made.

The remainder of the paper is organized as follows.
Section~\ref{sec:concept} presents the voltage sensor design and the basic circuit model that is ultimately used to extract the potential difference between a pair of conductors from a {single-contact} measurement.  In section~\ref{sec:analysis}, the equivalent circuit model is analyzed and we identify two design criteria that must be satisfied in order to make reliable voltage measurements.     The prototype sensor that was built and experimentally tested is discussed in section~\ref{sec:prototype}.  Section~\ref{sec:results} presents the key experimental results and section~\ref{sec:HV} describes a simple modification of the design that makes the voltage sensor suitable for high-voltage applications without compromising its performance.  Finally, the main conclusions are summarized in section~\ref{sec:conclusions}.

\section{Concept \& Model}\label{sec:concept}

This work is restricted to single-phase applications in which a pair of conductors, one neutral and one live, are suspended above earth ground. It is assumed that there is a good electrical connection between the neutral conductor and ground.  {Figure~\ref{fig:Fig1} shows two transmission line geometries that commonly exist in residential areas.  
\begin{figure}[t] 
\centering
\includegraphics[width=\columnwidth]{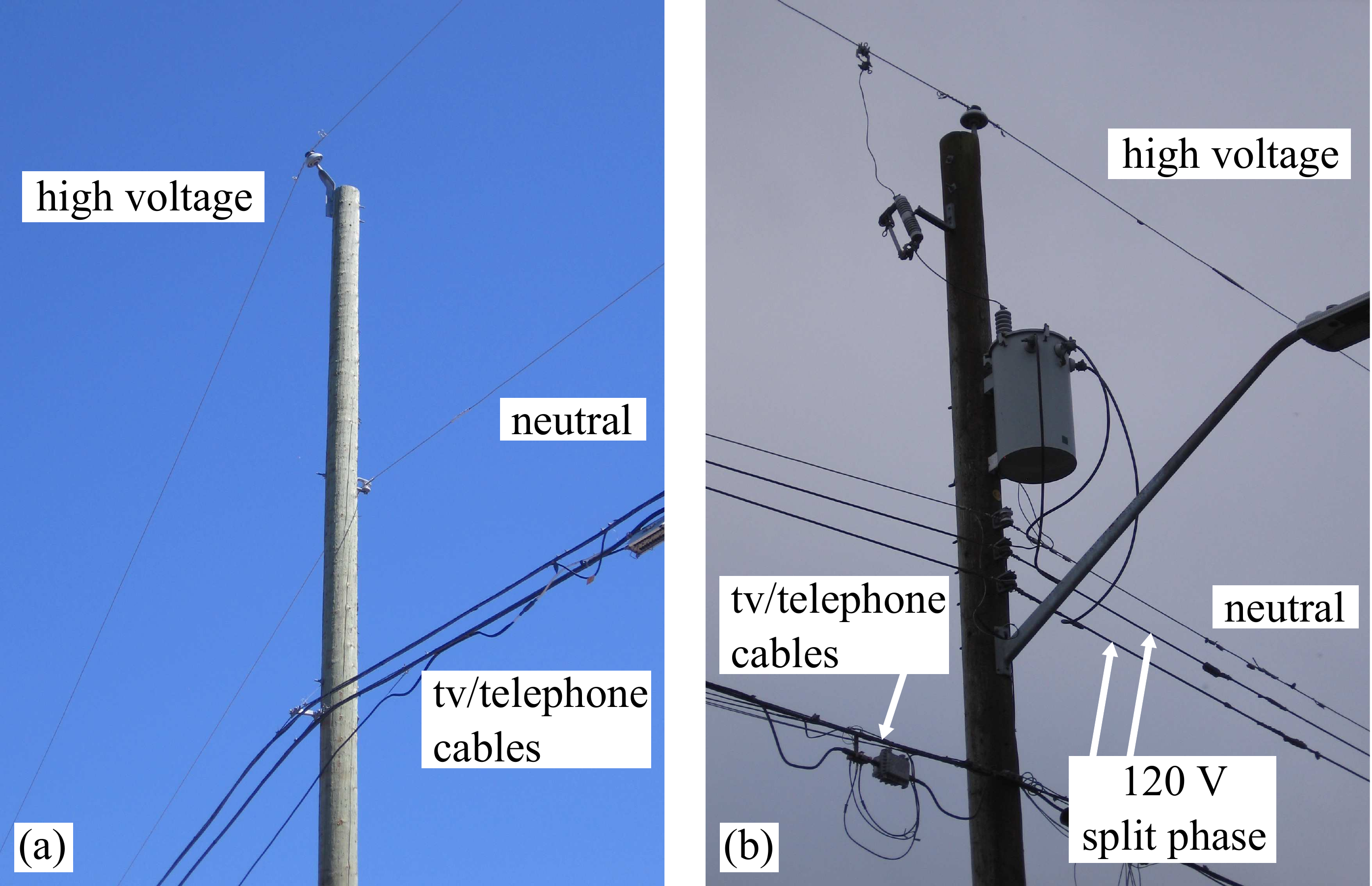}
\caption{{Two transmission line geometries that can be effectively monitored by the single-contact voltage sensor.  (a) A high-voltage conductor suspended above a neutral conductor.  (b) A high-voltage conductor suspended above a neutral conductor and two low-voltage split-phase conductors.}}
\label{fig:Fig1}
\end{figure}
In both of these cases the single-phase sensor described in this paper could be used to actively monitor the absolute electric potential difference between the neutral and the high-voltage lines.  In Fig.~\ref{fig:Fig1}(a), a high-voltage conductor is suspended above a neutral conductor.  In Fig.~\ref{fig:Fig1}(b), a center-tapped transformer steps down the potential of the topmost high-voltage line and supplies a neutral conductor and two 120~Vrms split-phase (antiphase) conductors below.  In the second case, a voltage sensor suspended from the high-voltage conductor will operate effectively as a single-phase sensor because, at the location of the high-voltage conductor, the net electric field due to the split-phase signals will almost exactly cancel.  In addition, the magnitude of the electric fields due to the 120~Vrms lines will be small compared to that of the high-voltage line which typically operates between 6 and 7~kVrms.}

{A schematic diagram highlighting the fundamental design aspects of the single-point contact voltage sensor is shown in Fig.~\ref{fig:Fig2}. 
\begin{figure}[t] 
\centering
\includegraphics[width=\columnwidth]{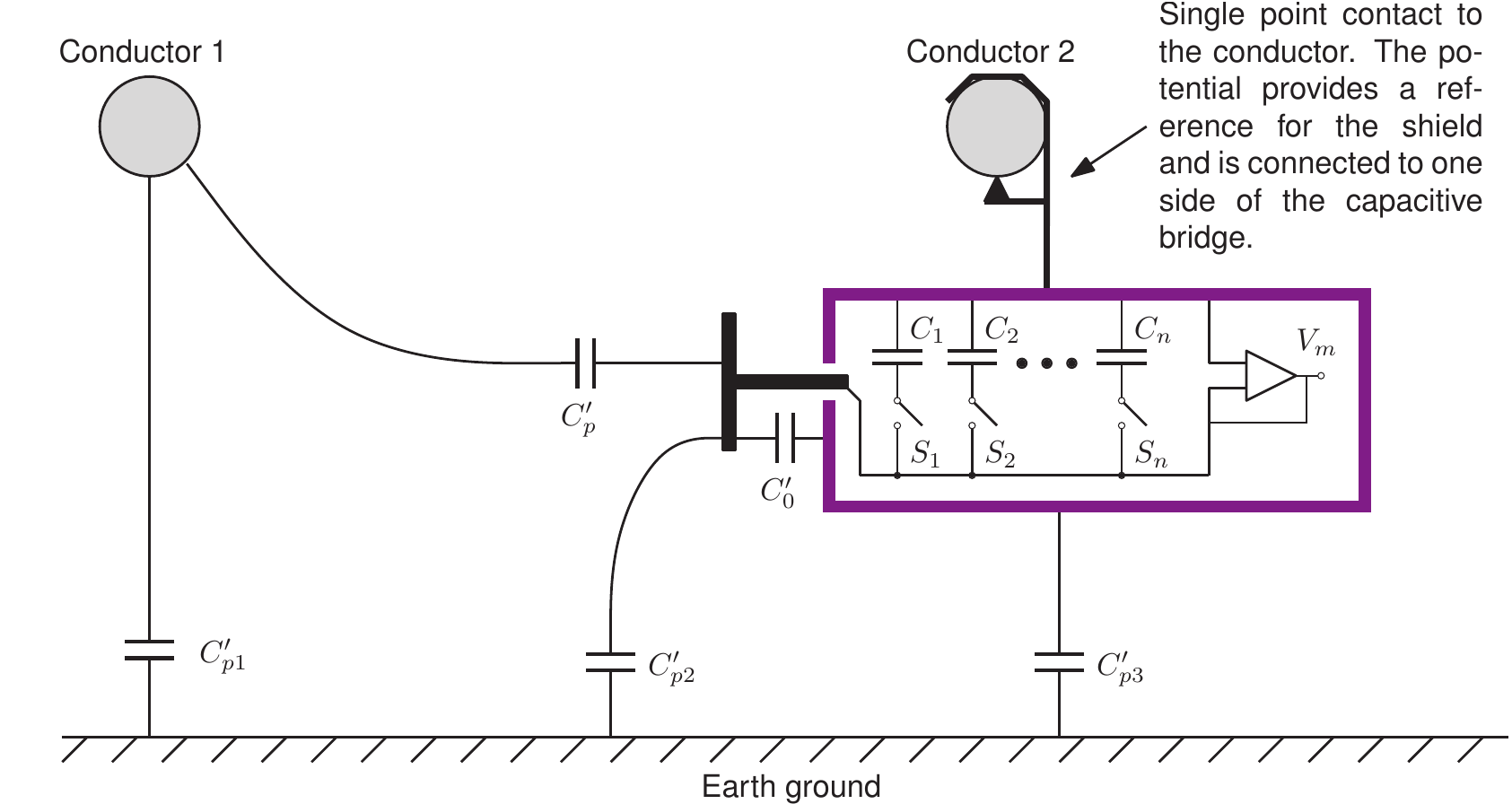}
\caption{{Schematic diagram of the single-contact voltage sensor.  A capacitive sensor plate passes through the main body of the sensor without making electrical contact to it.  The sensor body provides electromagnetic shielding and houses a tunable capacitor bank, a high-impedance buffer, and voltmeter (not shown).}}
\label{fig:Fig2}
\end{figure}
A capacitive sensor plate passes through, but is electrically isolated from, the main body of the sensor. The main body of the sensor is electrically and mechanically attached to conductor 2 which can be either the neutral or the live conductor}.  The body of the sensor, shown as the thick purple outline in the figure, is electrically conducting and provides electromagnetic shielding for the detection circuitry inside the enclosure. The shielding is crucial and ensures that the total capacitance between the sensor plate and the body is controlled and predictable. The total (net) capacitance between the sensor and body is called $C_0$. A second effective capacitance called $C_p$ is the net capacitance between the sensor plate and the non-contacted conductor, conductor 1. Together, $C_0$ and $C_p$ form a capacitive divider between the two conductors. As discussed in detail in the next section, the sensitivity of the voltage sensor is determined by the ratio of  $C_p$ over $C_0$ and therefore significant consideration must be given to controlling these capacitances in the sensor design.

There can be multiple contributions to the net capacitances of $C_p$ and $C_0$ and the contributions depend whether conductor~2 is neutral or live.  With reference to Fig.~\ref{fig:Fig2}, consider the case where conductor 2 is the neutral conductor.  In this case, $C_{p3}'$ is shorted by the neutral-earth connection, $C_{p2}'$ is in parallel with $C_0'$, and $C_{p1}'$ is the capacitance between the two conductors which has no effect on the voltage divider formed by $C_p$ and $C_0$. Therefore, in this configuration, $C_p=C_p^\prime$ and $C_0=C_0^\prime+C_{p2}^\prime$. On the other hand, if conductor 2 is the live conductor, $C_p=C_p^\prime+C_{p2}^\prime$ and $C_0=C_0^\prime$.  In this case, $C_{p1}^\prime$ is short-circuited by the neutral-earth connection and $C_{p3}^\prime$ is the effective capacitance between the two conductors which has no effect on the capacitive voltage divider.  {As a result, the conductor that the sensor is suspended from determines whether $C_p$ or $C_0$ is increased by an amount $C_{p2}^\prime$.}  Finally, housed inside the body of the sensor is a tunable capacitor bank placed in parallel with $C_0$.  The voltage $V_m$ is measured across $C_0$ in parallel with the net capacitance of the capacitor bank. The voltage measurement is buffered  and then connected to an analog-to-digital (A/D) converter, standard  rms ac voltmeter, {or oscilloscope}. The buffer must have very high input impedance and is essential to prevent loading the capacitive divider network.

{Outside of residential areas, three-phase transmission lines are commonplace.  This geometry is complicated by the fact that, in addition to the capacitance to earth ground, one must now consider the stray capacitances between the sensor and the two conductors that it is not in contact with.  Three-phase applications will be a topic of future investigations.}

\section{Voltage Sensor Analysis}\label{sec:analysis}
\begin{figure}[t] 
\centering
\begin{tabular}{l}
\includegraphics[width= 8.75 cm]{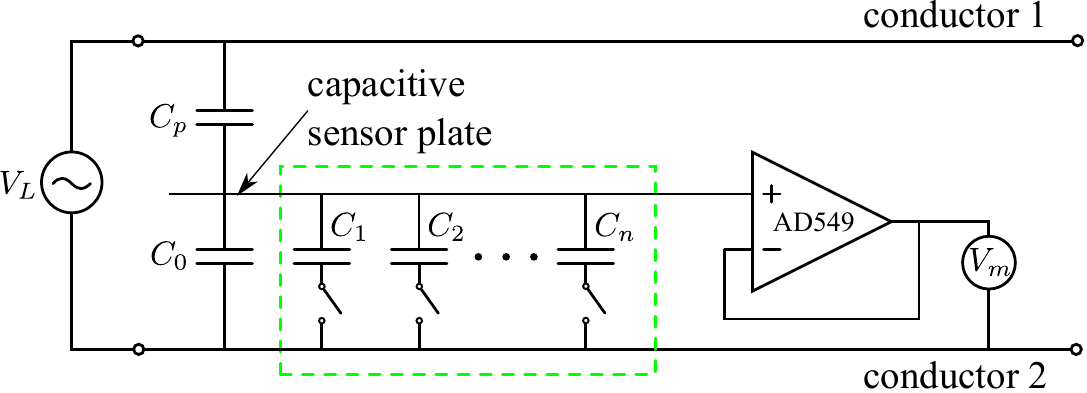}\\(a)\\~\\
\includegraphics[width= 8.75 cm]{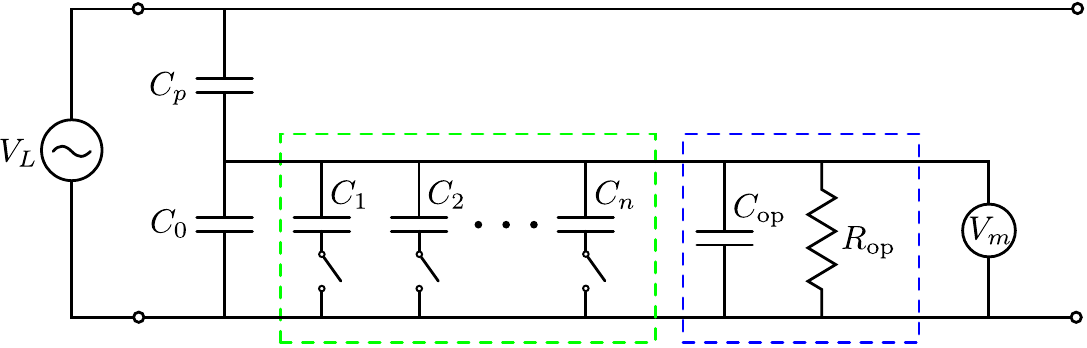}\\(b)
\end{tabular}
\caption{(a) Basic design of the single-contact voltage sensor.  A metallic sensor plate is used to create a capacitive divider between the two conductors of interest.  A switchable capacitor bank, shown in the dashed green rectangle, is placed between the sensor plate and the near conductor. A high-impedance buffer is placed between the sensor plate and the near conductor and its output voltage $V_m$ is measured. (b) The input impedance of the buffer is added in parallel to $C_0$ and the bank capacitance.}
\label{fig:Fig3}
\end{figure}

The equivalent circuit model of the single-contact voltage sensor is shown in Fig.~\ref{fig:Fig3}.  
In general, the input impedance of the op-amp must be taken into account when determining the final voltage $V_m$ measured at the op-amp output. {With the sensor attached to the neutral conductor, the voltage at the op-amp output is given by:
\begin{align}
V_m&=V_L\left[\frac{j\omega C_p}{R_\mathrm{op}^{-1}+j\omega\left(C_0^{\prime \prime}+C_b+C_p\right)}\right]\label{eq:1}\\
&=V_L\left[\frac{\omega^2 C_p\left(C_0^{\prime \prime}+C_b+C_p\right)+j\omega C_pR_\mathrm{op}^{-1}}{R_\mathrm{op}^{-2}+\omega^2\left(C_0^{\prime \prime}+C_b+C_p.\right)^2}\right]
\label{eq:2}
\end{align}
In these equations, $C_b=C_1+C_2+\dots$ is the net bank capacitance and:}
\begin{equation}
C_{0}^{\prime \prime}= C_0 + C_\mathrm{op} + C_s
\label{eq:3}
\end{equation}
where $C_0$  is the intrinsic capacitance between the sensor plate and the housing (defined earlier in section \ref{sec:concept}), $C_\mathrm{op}$ is the effective input capacitance of the op-amp,  and $C_s$ is any additional stray capacitance from the layout of the circuit inside the enclosure. We note that if the sensor is suspended from the live conductor, the expression for $V_m$ given in Eqs.~(\ref{eq:1}) and (\ref{eq:2}) will change only by an overall negative sign.  

{The frequency response of $V_m$ is that of a high-pass filter with a corner frequency of $f_c=\left[2\pi R_\mathrm{op}\left(C_0^{\prime \prime}+C_b+C_p\right)\right]^{-1}$.  The op-amp used as a buffer is an Analog Devices AD549L which has a nominal input resistance $R_\mathrm{op}$ of $10^{15}~\Omega$ in parallel with a nominal input capacitance $C_\mathrm{op}$ of $0.8$~pF~\cite{2008_Analog_Devices_AD549}.  Provided that the voltage sensor is operated at frequencies much greater than $f_c$, $R_\mathrm{op}$ makes negligible contribution to $V_m$.  Under these conditions, Eqs.~(\ref{eq:1}) and (\ref{eq:2}) simplify to:
\begin{equation}
V_m\approx V_L\left(\frac{C_p}{C_0^{\prime \prime}+C_b+C_p}\right)
\label{eq:4}
\end{equation} 
such that $V_m$ and $V_L$ are in phase and their relative amplitudes are determined by a ratio of capacitances.  Section~\ref{sec:results} will show that, for typical values of $C_0^{\prime \prime}$, $C_p$, and $C_b$, $f_c$ is expected to be less than 1~mHz.  The high-frequency cut-off of the voltage sensor is set by the gain-bandwidth product of the op-amp which is 1~MHz.  As a result, the voltage sensor is well suited to transmission line applications where the fundamental frequency (50 or 60~Hz) and harmonics fall well within the bandwidth of the design.} 

Equation~(\ref{eq:4}) also clearly shows that $V_m/V_L$ is greatest when $C_p$ is large and $C_0^{\prime \prime}$ is small and {this} explains why the voltage sensor sensitivity is enhanced when it is suspended from the live conductor.  Note that if one is only interested in relative voltage measurements, then $C_b$ can be set to any arbitrary value and the precise values of $C_0^{\prime \prime}$ and $C_p$ need not be known.  Furthermore, this sensor will reliably reproduce voltage waveforms and, if it is used in combination with a Rogowski coil, the power factor of the transmission line can be measured.   

In practice, absolute values of $V_L$ are extracted from measurements of $V_m$ as a function of $C_b$ and, although Eq.~(\ref{eq:4}) is very simple, there are important subtleties that must be considered in order to obtain reliable results.  First, if $V_{L}$, $C_p$, and $C_0^{\prime \prime}$ are all treated as unknowns, then the desired line voltage $V_L$ cannot be uniquely determined from measurements of $V_m$ as a function of $C_b$.  Any set of parameters $\{V_L,C_p, C_0^{\prime \prime}\}$ that maintains the correct values of the product $V_LC_p$ and the sum $C_0^{\prime \prime}+C_p$ will produce identical least-squares fits of Eq.~(\ref{eq:4}) to a set of $V_m$ versus $C_b$ measurements.  To extract absolute values of $V_L$ requires that {either} $C_0^{\prime \prime}$ or $C_p$ be known.  It is far easier to obtain reliable estimates of $C_0^{\prime \prime}$ via a factory calibration because its value will be larger than $C_p$ and it will be less dependent on the geometry of the conductors. (See section~\ref{sec:results}.)  

The second crucial requirement for accurate voltage measurements is that $C_p$ must {be large enough to contribute significantly} to the denominator of Eq.~(\ref{eq:4}) for at least some of the $C_b$ values.  If this condition is not satisfied, then \mbox{$V_m\approx V_LC_p/\left(C_0^{\prime \prime}+C_b\right)$} and one is only able to determine the product $V_LC_p$.  Requiring $C_0^{\prime \prime}$ to be on the same order of $C_p$ presents a major design challenge because, as will be shown in section~\ref{sec:results}, $C_p$ can be 1~pF or less for typical sensor plate/transmission line geometries. Therefore, it is critical to design a capacitive sensor that maximizes the ratio $C_p/(C_0+C_p)$ and to use a circuit layout that minimizes stray capacitance contributions to $C_0^{\prime \prime}$.    

\section{Prototype \& Circuit Layout}\label{sec:prototype}
\begin{figure}[t] 
\centering
\includegraphics[width=0.8\columnwidth]{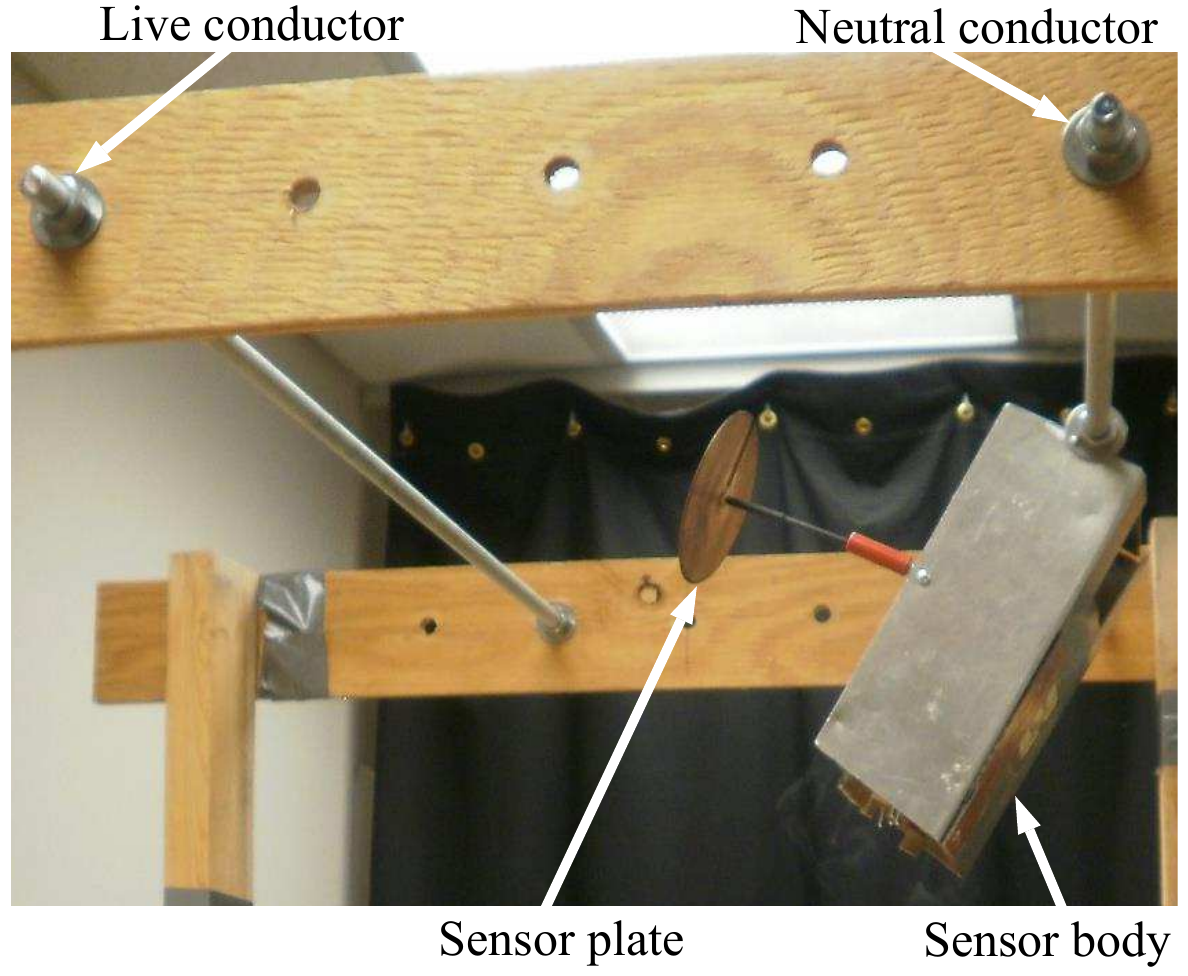} 
\caption{Digital photograph of the voltage sensor anchored to the neutral conductor of a short section of transmission line.  The capacitive sensor plate is suspended between the two conductors.  The buffer circuit is contained within the shielded sensor body.}
\label{fig:Fig4}
\end{figure}

Figure~\ref{fig:Fig4} shows a photograph of the prototype voltage sensor that was constructed and experimentally tested.  
The main body of the sensor is made of aluminum and anchored to the neutral conductor of a short section of transmission line constructed from half-inch diameter threaded steel rod mounted in a wooden frame.  A rigid conducting rod that passes through, but is electrically isolated from the sensor body, supports the copper capacitive sensor plate.  

Several capacitive sensor designs were explored both experimentally and via numerical simulations before settling on a plate suspended from a thin rod.  In the limit that the live conductor is very far from the capacitive sensor, the self-capacitance of the sensor is expected to determine the value of $C_p$.  For example, a spherical sensor that has the same radius as the disk would have a self-capacitance that is larger by a factor of $\pi/2$.  However, as will be shown in the next section, this modest gain in $C_p$ is offset by an even larger gain in $C_0$ resulting in an overall decrease in the ratio $C_p/(C_0+C_p)$.   The $C_p/(C_0+C_p)$ ratio is maximized by the thin disk sensor simply because this geometry minimizes $C_0$ by keeping the sensor plate as far as possible from the main body of the sensor.                

\begin{figure}[t] 
\centering
\includegraphics[width=\columnwidth]{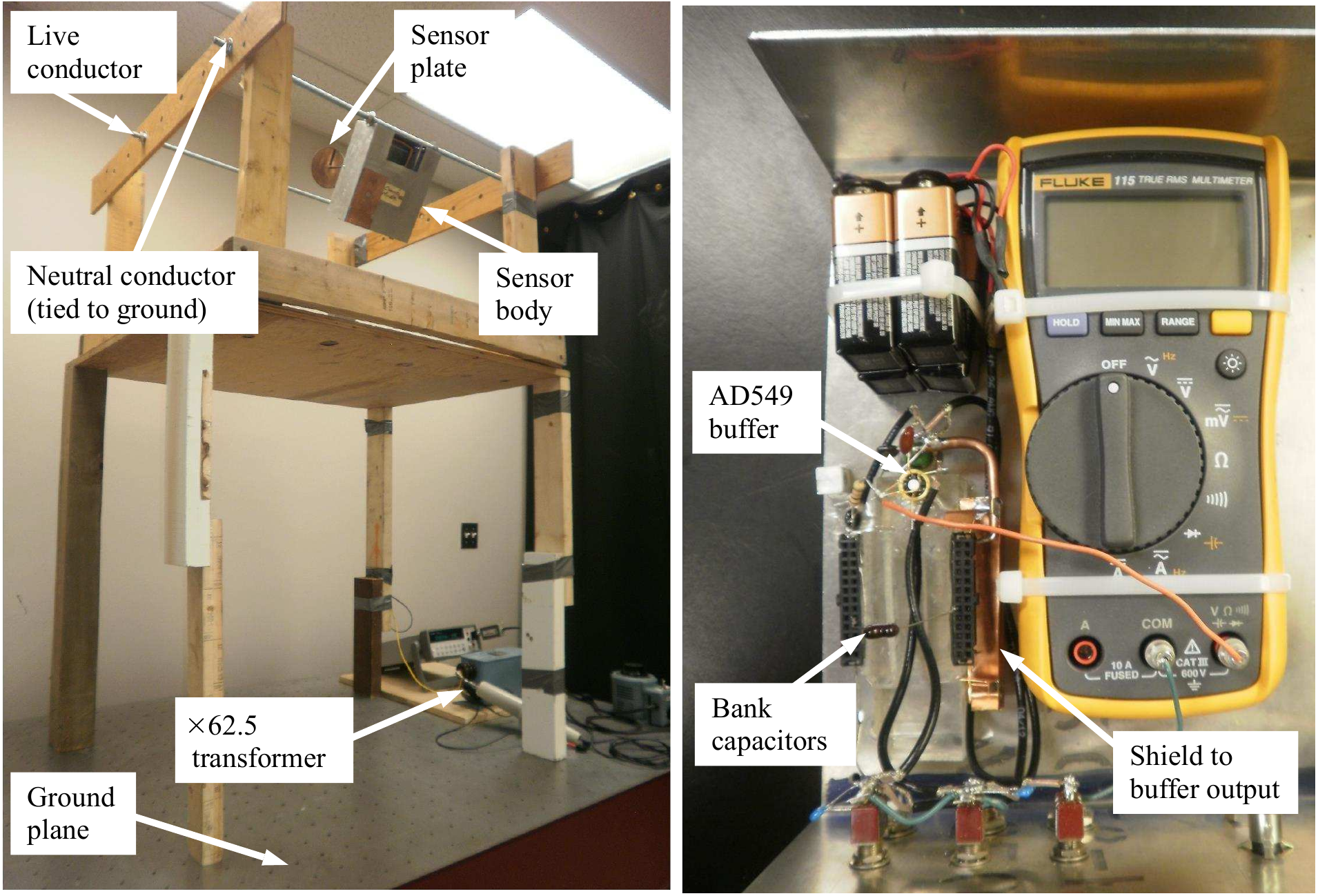} 
\caption{The left photograph shows the experimental apparatus consisting of the voltage sensor and a short section of transmission line with adjustable conductor spacing and height that is powered using a variac and a high-voltage transformer.  The entire apparatus is placed on a grounded optical table.  The photograph on the right shows the buffer circuit, capacitor bank, and a hand-held digital multimeter contained inside the sensor body.}
\label{fig:Fig5}
\end{figure}

{The sensor electronics are isolated from stray fields by enclosing them inside a conducting box that serves as the main body of the voltage sensor.}  The circuit layout is shown in Fig.~\ref{fig:Fig5}.
The op-amp circuit is powered by $\pm 18$~V supplies using 9~V batteries in series and the op-amp output is measured by a Fluke~115 digital multimeter {or, if one wishes to observe the waveforms, an oscilloscope}.  In a commercial application, the op-amp output would be captured by an A/D converter and then wirelessly transmitted to a {remote} receiver.  {The A/D converter can sample the op-amp output at a rate much greater than 60~Hz such that the transmitted signal reliably reproduces a scaled version of the true waveform of the transmission line signal.}  

The most important consideration {for the} buffer and capacitor bank circuit is layout.   An optimized layout that (a) minimizes stray capacitance $C_s$ between the signal line originating from the sensor plate and the main body of the sensor and (b) preserves the very high input resistance of the op-amp must be used.  Many useful guidelines are given in Refs.~\cite{2008_Analog_Devices_AD549} and \cite{1997_Analog_Devices_AD515A}.  First, we have avoided using circuit boards which can add both unwanted capacitance and a finite resistance through the dielectric insulator.  As shown in the {Fig.~\ref{fig:Fig5}}, all electrical connections to the op-amp have been made by soldering leads directly to the op-amp pins. Additionally, connections between the sensor plate and the buffer input are shielded with coaxial cables whose outer conductors are tied to the op-amp output.  This eliminates leakage currents through the coaxial cable dielectric, guards against outside interference, and eliminates stray capacitance between the op-amp input signal line and the surrounding neutral enclosure of the voltage sensor~\cite{1997_Analog_Devices_AD515A}.  Also visible in Fig.~\ref{fig:Fig5}, are the receptacles for the bank capacitors.  The left-hand side receptacle is tied to the sensor enclosure and the right-hand side to the buffer input.  A copper foil shield is held at the output potential of the buffer, once again, to minimize unwanted stray contributions to $C_0^{\prime \prime}$.   

To get an estimate of the combined contributions of $C_\mathrm{op}$ and the stray capacitance $C_s$ to $C_0^{\prime \prime}$, the sensor disk and its support rod were removed and the bank capacitance was set to zero.  A known calibration capacitor $C_\mathrm{cal}$ was placed between the buffer input and the far (live) conductor. A calibration voltage $V_\mathrm{cal}$ was then applied across the neutral and live conductors. Under these conditions, the op-amp output is:
\begin{equation}
V_m=\frac{V_\mathrm{cal}C_\mathrm{cal}}{C_s+C_\mathrm{op}+C_\mathrm{cal}}.\label{eq:5}
\end{equation}          
Rearranging this expression gives the desired result: \mbox{$C_s+C_\mathrm{op}=C_\mathrm{cal}\left(V_\mathrm{cal}/V_m-1\right)$}. Using this method, $C_s+C_\mathrm{op}$ was measured to be $5.97\pm 0.06$~pF where the uncertainty is associated with the 1\% tolerance  of the silver mica calibration capacitor, $C_\mathrm{cal}$.

We note that in the prototype sensor presented here, the bank capacitance is changed manually by adding and removing individual silver mica capacitors.  In a practical sensor, the bank capacitance values would be changed remotely perhaps using MOSFET or MEMS switches in series with each individual bank capacitor.

\section{Experimental and Simulation Results}\label{sec:results}

In this section, simulation and experimental results are presented to evaluate the performance of the single-point contact voltage sensor.  {Various capacitive sensor designs were evaluated to determine the sensing plate geometry that maximizes the sensitivity of the sensor.} Next, the dependencies of $C_p$ and $C_0$ on the spacing between conductors and the conductor height above the ground plane were investigated using both experimental measurements and simulations.  The sensor output voltage $V_m$ was also tested with the transmission line terminated with different loads (open, resistive, capacitive, and inductive) and shown to be independent of the line current.  {The bandwidth of the sensor was then experimentally measured and, as expected, shown to be flat over the frequency range of interest.}  Finally, and most importantly, the prototype sensor was used to extract the line voltage $V_L$ from measurements of $V_m$ as a function of the bank capacitance $C_b$.  Our measurements show that the voltage sensor operates reliably over a wide range of line voltages.

\subsection{Simulation Model \& Sensor Shape}\label{sec:sim}

\begin{figure}[t] 
\centering
\includegraphics[width=4.5 cm]{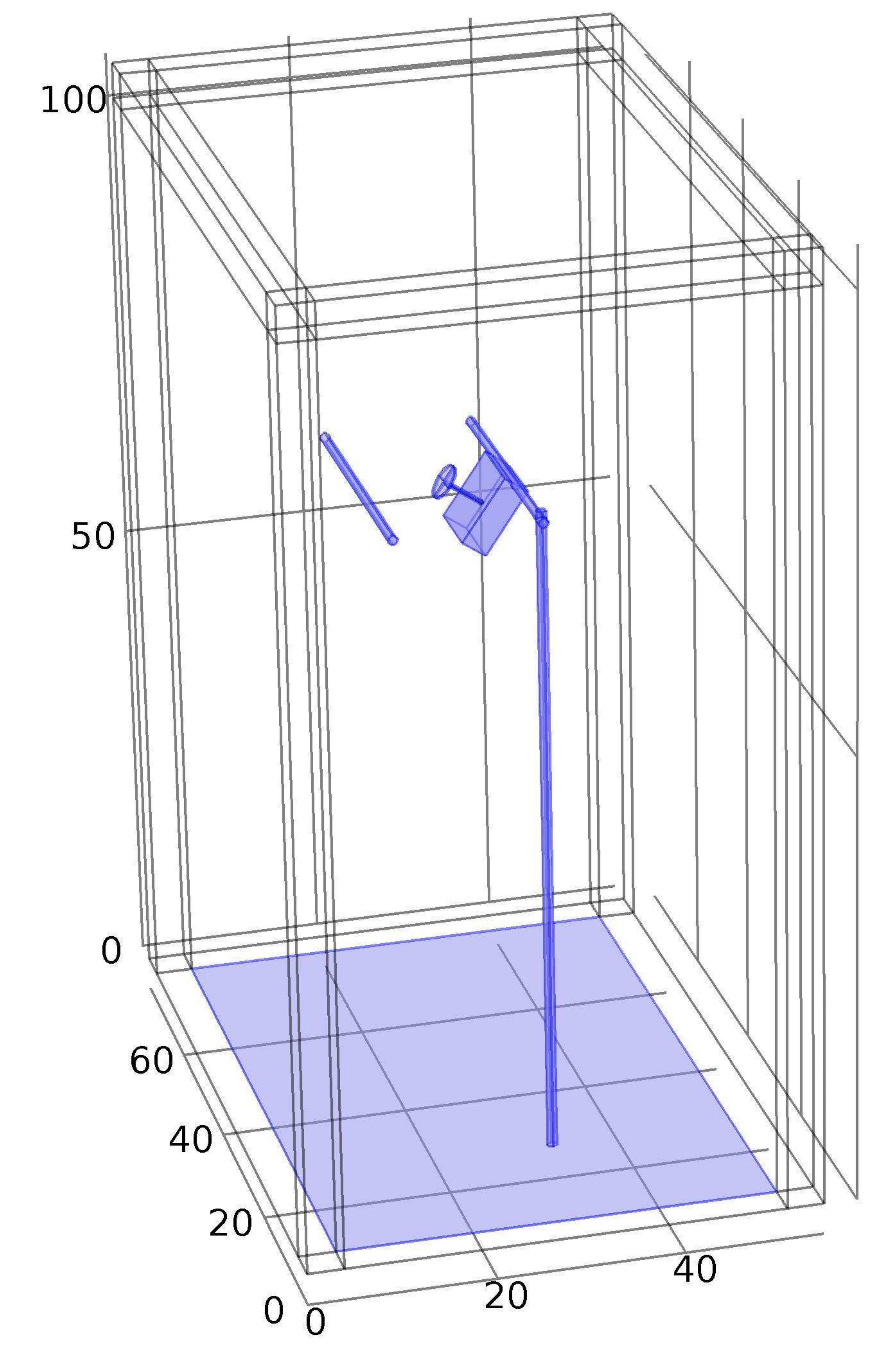}
\caption{An example of the simulation model used for evaluating the sensor.  In this model, the sensor is suspended from the neutral conductor which is tied to the ground plane below the conductors.  The scales shown on the $x$-, $y$-, and $z$-axes are in inches.}
\label{fig:Fig6}
\end{figure}

An electrostatic model of the experimental test bed was created in a finite element electromagnetic simulation tool (COMSOL version 4.3b). Figure~\ref{fig:Fig6} shows a scale model of the sensor geometry used in the simulations.   In this figure, the voltage sensor is shown suspended from the neutral conductor. The neutral conductor is electrically connected to the ground plane by a long vertical conductor. The ground plane mimics earth ground in a high-voltage transmission system. In the model, a capacitive sensor consisting  of  a thin disk  points towards the opposite high-voltage conductor.

The simulations in COMSOL are used to estimate the capacitances $C_0$ and $C_p$. A rectangular volume encloses the conductors, sensor, and ground plane. The volume determines the region over which the electric field is calculated in the simulation.  Since the volume in a practical application is an open domain, the boundary conditions for the top and sides of the bounding surface are defined as infinite domain surfaces which means field lines can extend beyond the volume and are not artificially constrained to lie within the volume. The ground plane is the bottom surface of the volume and  has dimensions of 1.77~m by 1.19~m; the dimensions   match the  size of the optical table which was used as a ground plane in the experiment. The distance between the conductors and the top of the volume was fixed at 76.2~cm. Later, sweep results are shown for different conductor heights above the ground plane. For the height sweep, the distance from the conductors to the top surface of the volume is held constant.

The capacitances $C_p$ and $C_0$ were extracted by setting the potential of conductors in the model. Since capacitance is independent of potential, a normalized potential of 1~V was selected.  The individual capacitances are then found by surface integrals (Guass' law) over either the neutral or live (high-voltage) conducting surfaces. Since the potential across the capacitances is 1~V, the capacitance is equal to the total charge enclosed within the surface ($C = Q$ for $V = 1$~V).  

As shown in Fig.~\ref{fig:Fig6}, the sensor plate extends out of an aluminum box which houses the sensor electronics. The box has dimensions of 20.5~cm tall, 19.5~cm wide, and 7.6~cm deep.  The box is tilted at an angle of $35^\circ$ to point the sensor face towards the non-contacting conductor. The sensor plate passes through the aluminum housing via a circular aperture. The aperture is offset to one side and is 9.5~cm from the top and 3.4~cm from the side. A clearance ring with a thickness of 2~mm is provided between the sensor rod and the housing. All conductors, including the aluminum housing,  are modeled as perfect electric conductors.

As discussed in section~\ref{sec:analysis} and shown by Eq.~(\ref{eq:4}), the sensitivity of the voltage sensor is determined by the ratio $C_p/(C_0+C_p)$.  Clearly, the most effective way to enhance {this ratio} is by moving the capacitive sensor away from the main body of the sensor (decrease $C_0$) and towards the opposite conductor (increase $C_p$). An objective of this work was to design a low-profile sensor and extending the sensor plate significantly from the body of the sensor was not considered to be of interest. Therefore, the extension of the sensor plate from the body of the sensor was constrained to a modest distance of 12.5~cm in this work. Also, in high-voltage applications, there are practical considerations, such as the risk of arcing,  that limit how close the capacitive sensor can extend towards the far conductor.  

Four different sensor geometries were studied using simulations. The designs are: 1) a thin disk, 2) a parabolic sensor, 3) a cone, and 4) a sphere. The  thin disk is defined by a  radius $R$ supported by a thin rod of radius $r$. The thin disk is shown in Fig.~\ref{fig:Fig6} and was used in the experimental design. The other sensor designs evaluated by simulation include a parabolic sensor whose shape is generated by revolving the parabola \mbox{$y=(R-r)(x/L)^2+r$} about the $x$-axis for \mbox{$0<x<L$}, a conical sensor whose shape is generated by revolving the straight line \mbox{$y=(R-r)(x/L)+r$} about the $x$-axis for \mbox{$0<x<L$}, and a spherical sensor of radius $R$ supported by a thin rod of radius $r$.  For all designs studied,  $R=4.8$~cm, $r=1.12$~mm, and $L=12.5$~cm.  The thin disk sensor had a thickness $t=1.6$~mm. Images of the sensor designs are shown in Table~\ref{tab:Tab1}; note that the parabolic and conical sensors are closed volumes with flat apertures.

\begin{table}[!t]
\renewcommand{\arraystretch}{1.3}
\caption{Simulated values of $C_p/(C_0+C_p)$ for different sensor plate shapes.  Results are given both for the voltage sensor suspended from the neutral conductor and from the live conductor. ($d=40~\mathrm{cm}$, $h=143~\mathrm{cm}$)}
\label{tab:Tab1}
\centering
\begin{tabular}{cccccc}
\hline\hline\\[-0.3cm]
\multicolumn{6}{c}{\emph{suspended from neutral conductor}}\\
\hline\\[-0.35cm]
\multicolumn{2}{c}{Sensor shape} & $C_p$ (pF) & $C_0$ (pF) & $\dfrac{C_p}{C_0+C_p}$ & $\dfrac{C_p}{C_0^{\prime \prime}+C_p}$\\[0.25 cm]
\hline\\[-0.2cm]
thin disk & \includegraphics[width=0.5 cm]{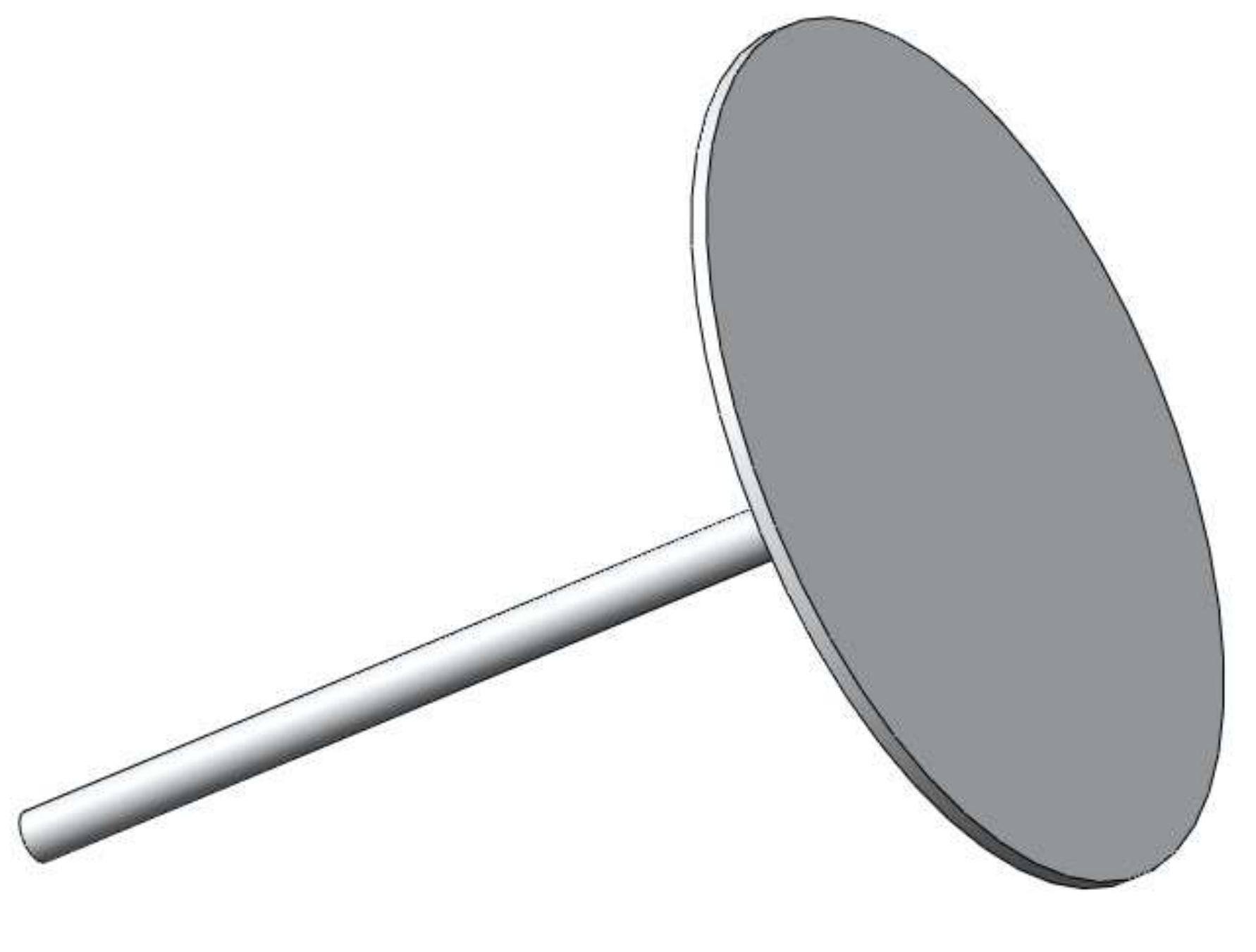} & 0.784 & 3.29 & 0.192 & 0.0781\\
parabolic & \includegraphics[width=0.5 cm]{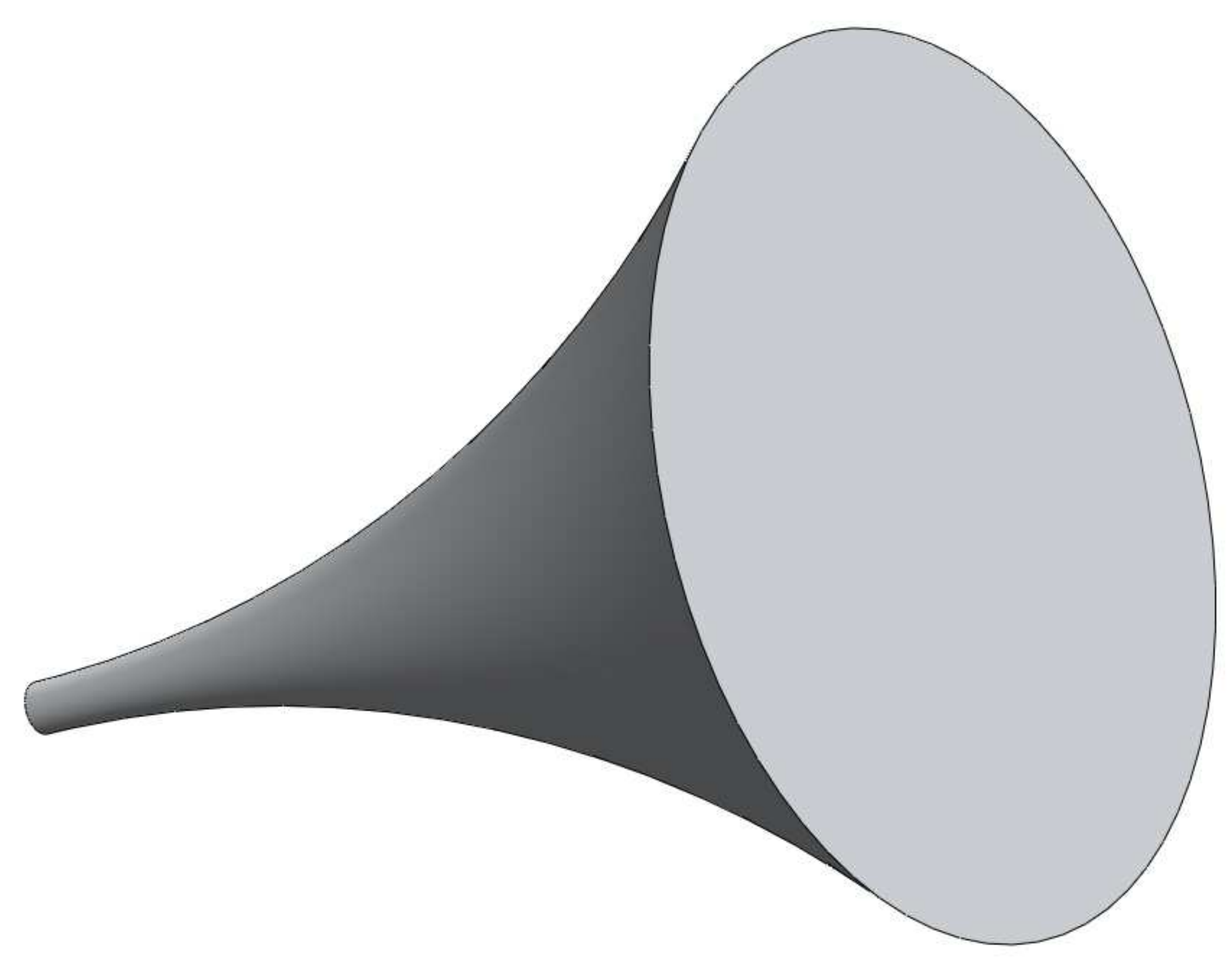} & 0.893 & 4.12 & 0.178 & 0.0813\\
conical & \includegraphics[width=0.5 cm]{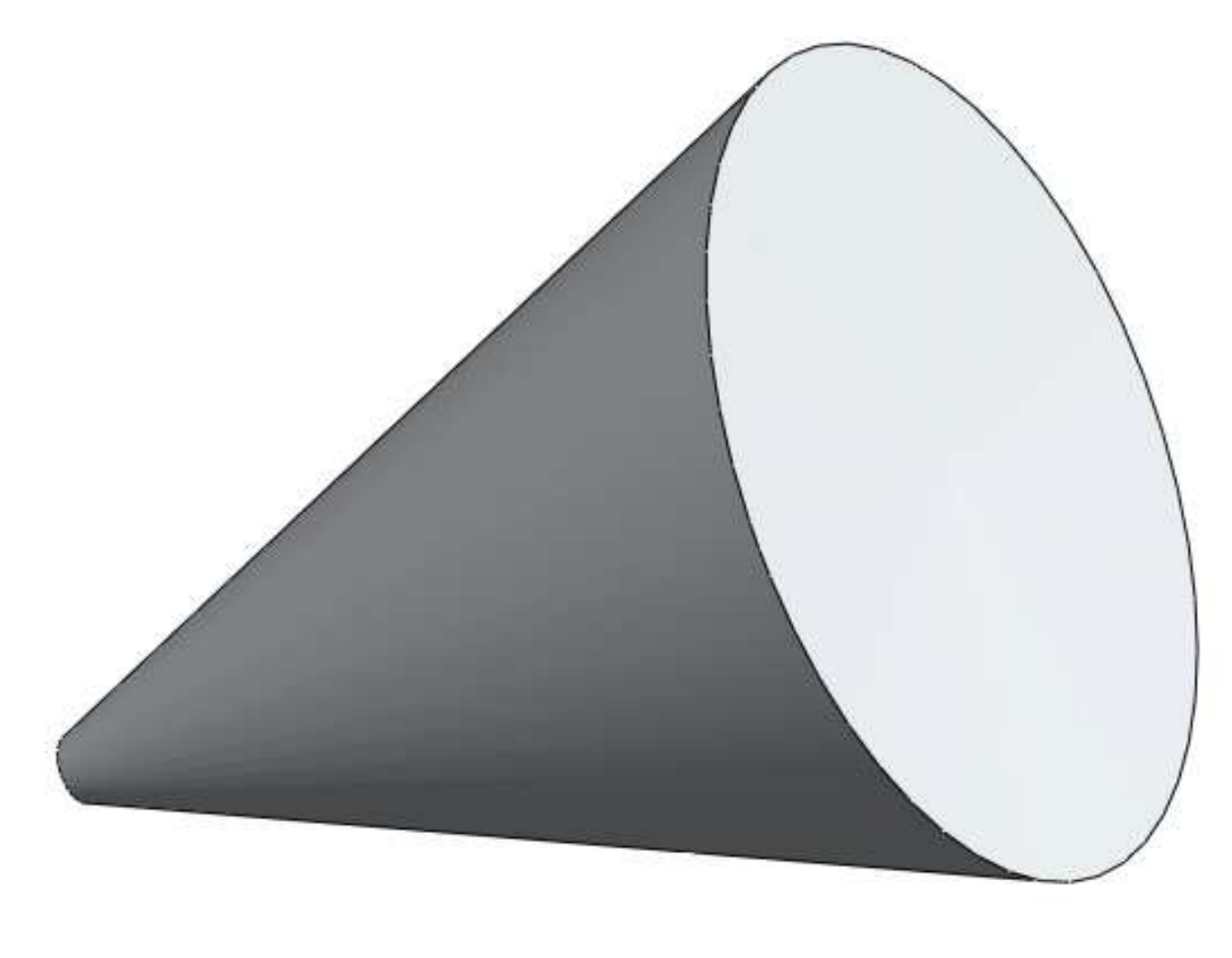} & 0.965 & 5.05 & 0.160 & 0.0805\\
spherical & \includegraphics[width=0.45 cm]{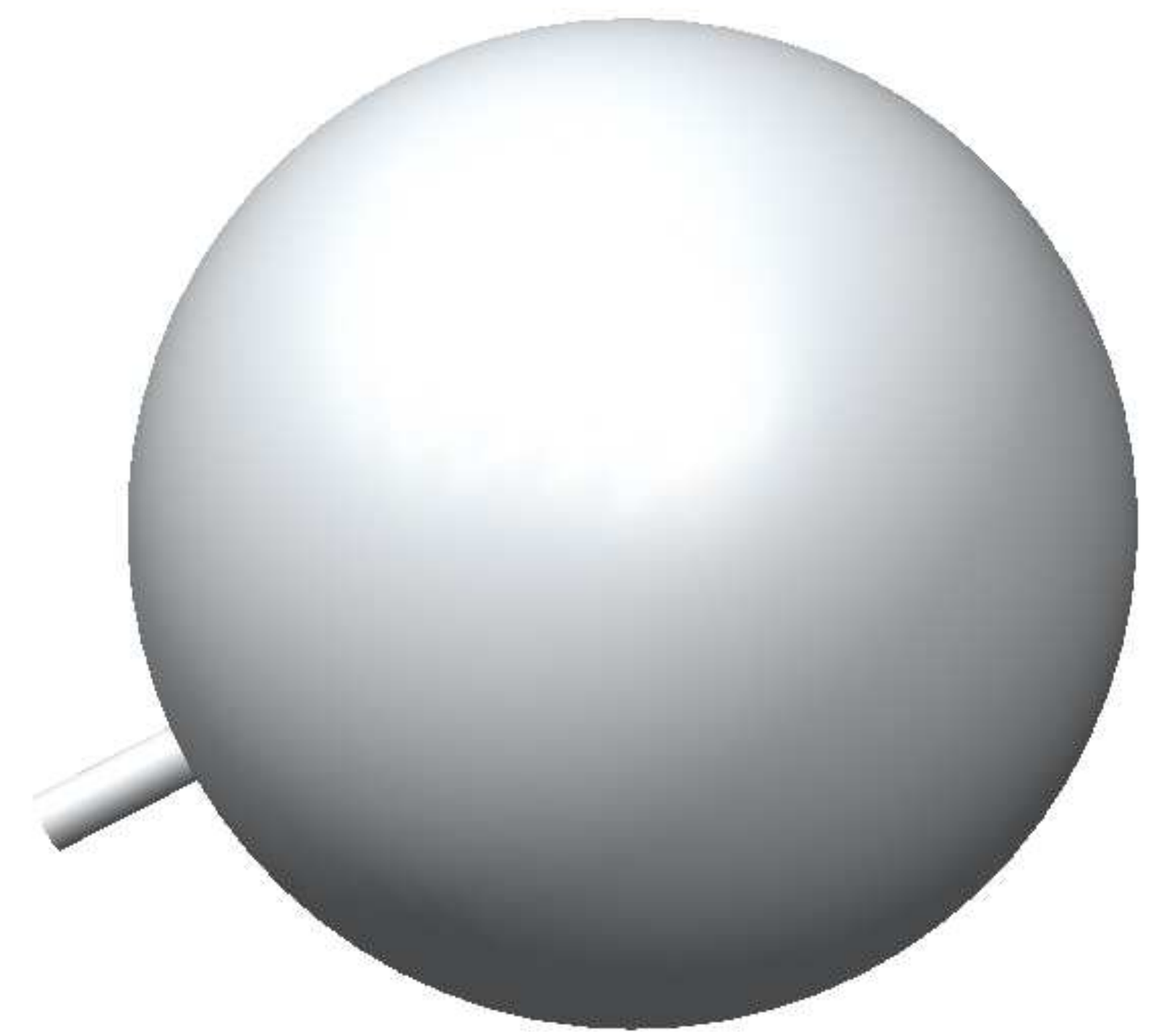} & 1.13 & 6.39 & 0.150 & 0.0838\\
\hline\hline\\[-0.3cm]
\multicolumn{6}{c}{\emph{suspended from live conductor}}\\
\hline\\[-0.35cm]
\multicolumn{2}{c}{Sensor shape} & $C_p$ (pF) & $C_0$ (pF) & $\dfrac{C_p}{C_0+C_p}$ & $\dfrac{C_p}{C_0^{\prime \prime}+C_p}$\\[0.25 cm]
\hline\\[-0.2cm]
thin disk & \includegraphics[width=0.5 cm]{BobowskiTab1a.pdf} & 1.33 & 2.73 & 0.328 & 0.133\\
parabolic & \includegraphics[width=0.5 cm]{BobowskiTab1b.pdf} & 1.56 & 3.51 & 0.308 & 0.141\\
conical & \includegraphics[width=0.5 cm]{BobowskiTab1c.pdf} & 1.62 & 4.27 & 0.275 & 0.137\\
spherical & \includegraphics[width=0.45 cm]{BobowskiTab1d.pdf} & 1.66 & 4.90 & 0.253 & 0.132\\
\hline
\end{tabular}
\end{table}

\begin{figure*}
\centering
\begin{tabular}{cc}
(a)\includegraphics[width=7.5 cm]{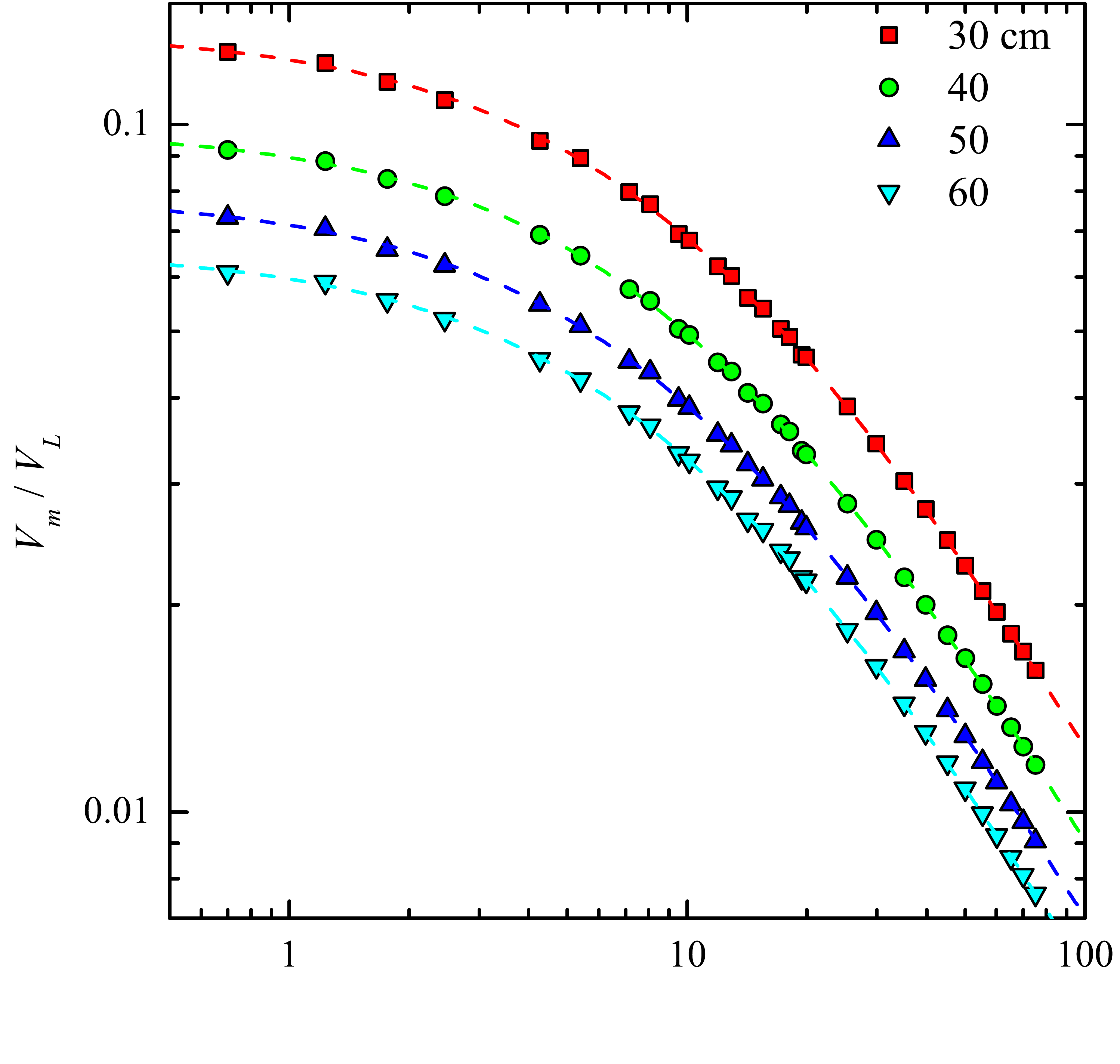} & (b)\includegraphics[width=7.5 cm]{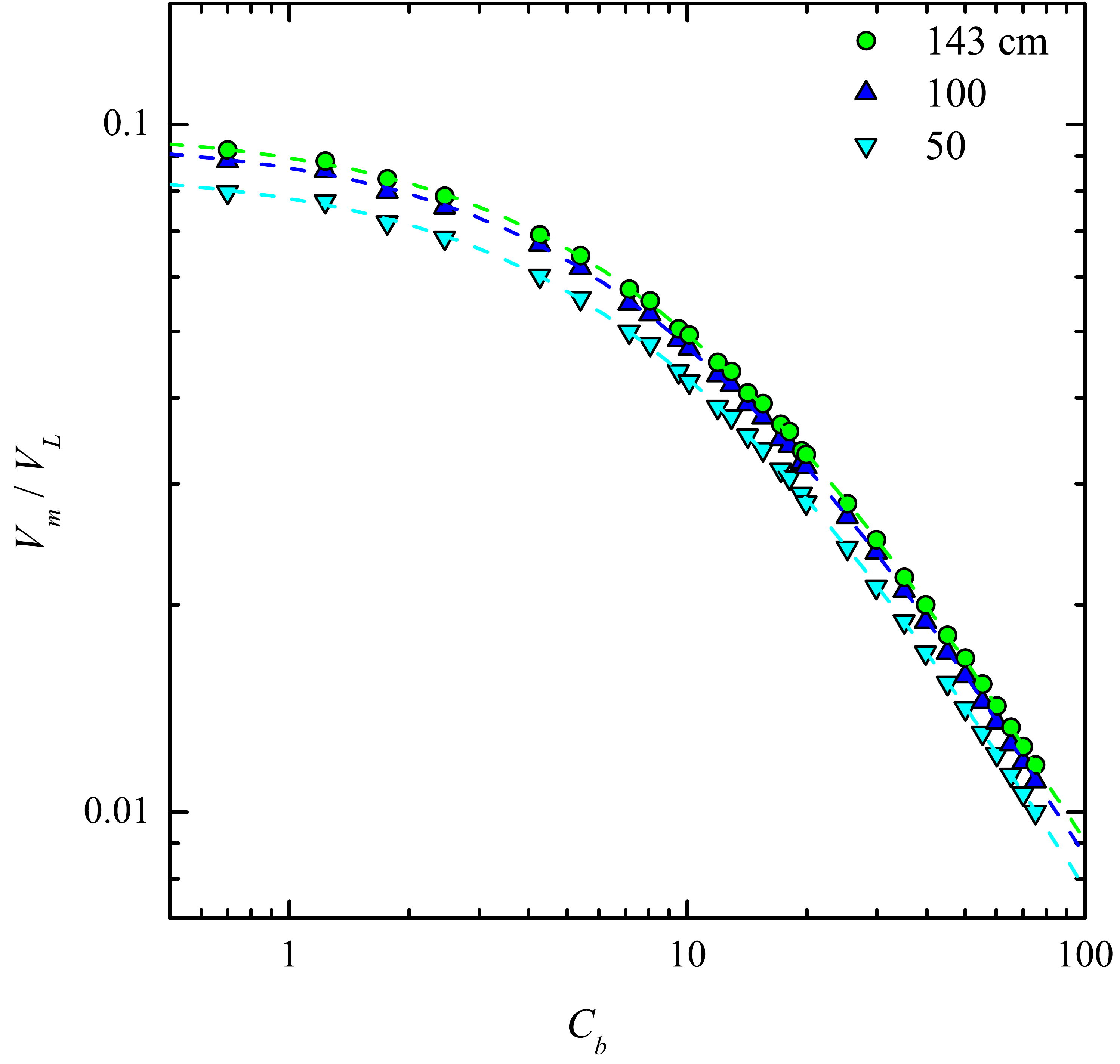}\\
~ & ~\\
(c)~\includegraphics[width=7. cm]{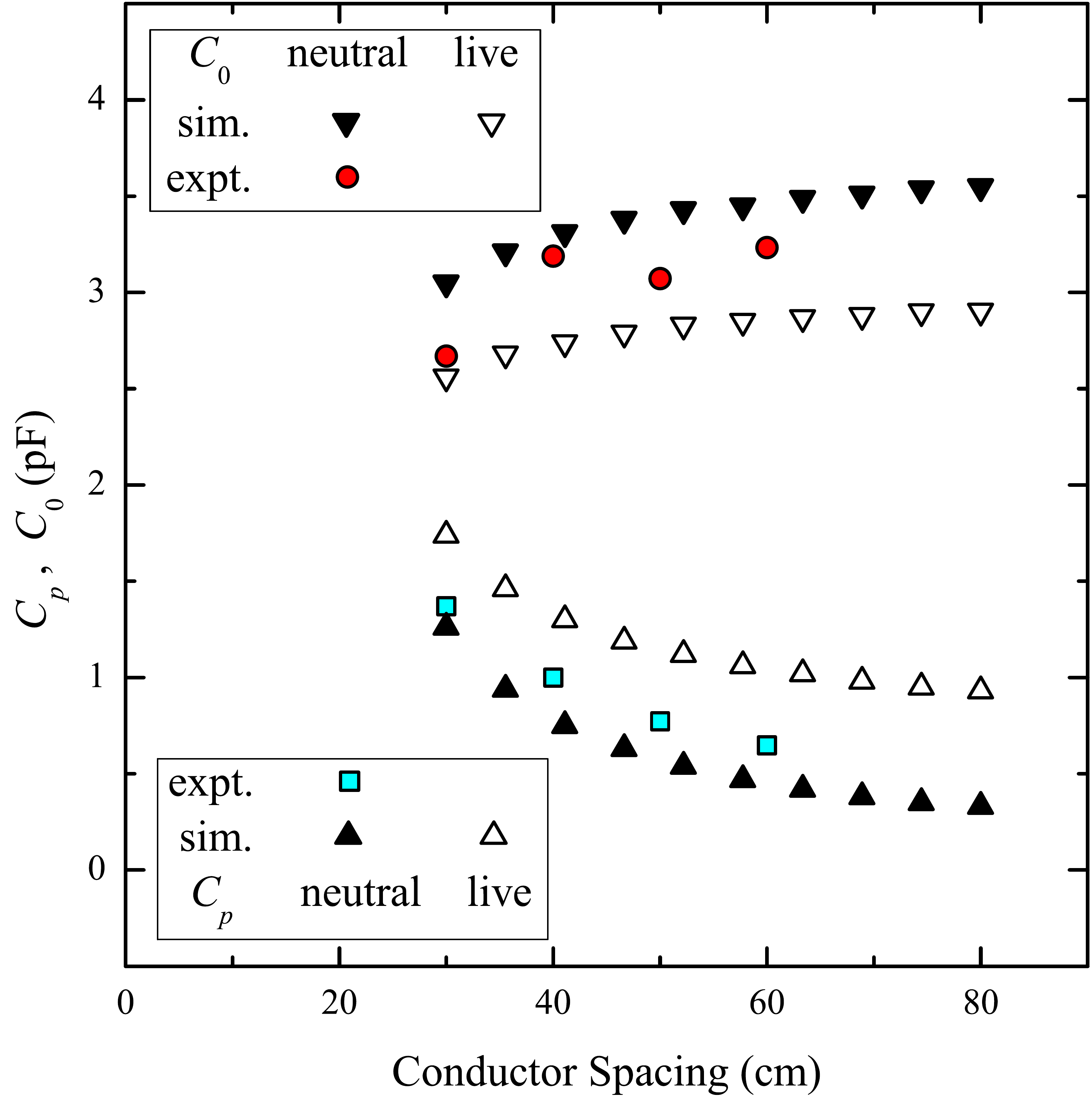} & (d)~\includegraphics[width=7. cm]{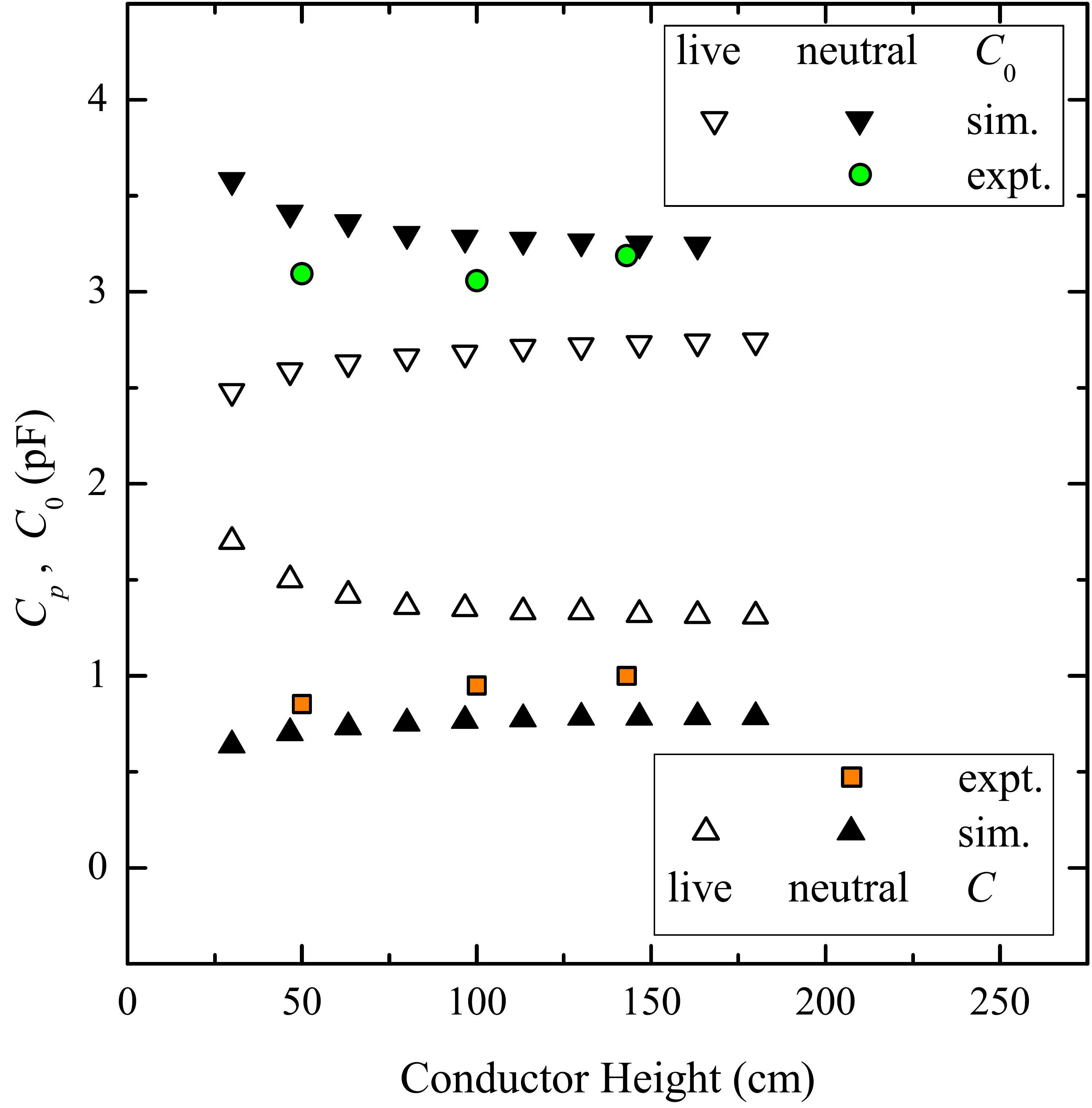}
\end{tabular}
\caption{{For all experimental measurements in this figure the voltage sensor was suspended from the neutral conductor.  (a) Measured $V_m/V_L$ as a function of $C_b$ for four different conductor spacings and $h=143$~cm with least-squares fits to Eq.~(\ref{eq:4}). (b) Measured $V_m/V_L$ as a function of $C_b$ for three different conductor heights above the ground plane and $d=40$~cm with least-squares fits to Eq.~(\ref{eq:4}). (c) Comparison of experimental and simulation results for $C_0$ and $C_p$ versus conductor spacing. (d) Comparison of experimental and simulation results for $C_0$ and $C_p$ versus conductor height.}}
\label{fig:Fig7}
\end{figure*}

The different sensor designs were evaluated systematically for a conductor spacing of $d=40$~cm and a conductor height above the ground plane of $h=143$~cm. Table~\ref{tab:Tab1} summarizes the simulation results for $C_p$, $C_0$, and the ratio $C_p/(C_0+C_p)$. The simulation results also include evaluating two different configurations: 1) where the sensor is attached to the neutral line and 2) when the sensor is attached to the  live conductor.  

The results show that the thin-disk capacitive sensor has the largest $C_p/(C_0+C_p)$ ratio and therefore the highest sensitivity.  Modifying the sensor geometry can produce a modest increase in $C_p$, but it is accompanied by an even larger increase in $C_0$.  For example, when suspended from the neutral conductor, a spherical sensor has a 44\% larger value of $C_p$ than the disk sensor; however, although a high value of $C_p$ is desirable, the spherical sensor has a value of $C_0$ which is 94\% larger than that of the disk sensor. Therefore, the spherical sensor has a capacitor ratio which results in less accuracy than the disk sensor. Of the four designs which were studied, the disk sensor is the best design in terms of maximizing the accuracy of the sensor.

Note that simulated values of $C_0$ do not include the additional contributions from the op-amp input capacitance and any stray capacitance due to the circuit layout that would be expected in a practical implementation of the voltage sensor design. The additional capacitances must be controlled to avoid increasing the net capacitance $C_0''$ (\mbox{$C_0^{\prime \prime}=C_0+C_\mathrm{op}+C_s$}) too much. Recall that a large value of $C_0''$ will start to limit the accuracy of the overall sensor. In this design,  $C_\mathrm{op}+C_s$ was measured to be approximately 6.0~pF, and several design iterations were made to minimize this value.

Another important observation from the results in  Table~\ref{tab:Tab1} is that voltage sensor sensitivity can be improved significantly by suspending it from the live conductor instead of the neutral conductor.  For all of the capacitive sensor shapes, and regardless of the value of $C_\mathrm{op}+C_s$, the sensitivity increases if the sensor is moved from the neutral to the live conductor.  The enhancement in sensitivity ranges from 58\% to 73\% and the improvement relates to the stray capacitor components which make up $C_0$ and $C_p$. The breakdown of these capacitances was explained earlier in sections~\ref{sec:concept} and \ref{sec:analysis} (see Fig.~\ref{fig:Fig2}).  As a result, when it is practical to do so, the voltage sensor should be suspended from the live conductor.

\subsection{Dependence of $C_p$ \& $C_0$ on Conductor Spacing and Height}\label{sec:expt}  

Results are now summarized for a configuration where the flat disk sensor is attached to the neutral conductor.  These experiments use the prototype sensor shown in Figs.~\ref{fig:Fig4} and \ref{fig:Fig5} with the line voltage $V_L$ held constant at 100~Vrms. In Fig.~\ref{fig:Fig7}(a), the sensor voltage is shown as a function of the bank capacitance $C_b$ for four different conductor spacings ranging from 30~cm to 60~cm. In Fig.~\ref{fig:Fig7}(b) a similar set of measurements are shown except in this test the conductor spacing was fixed at $d=40$~cm and measurements where made for three different conductor heights: $h=50$, 100, and 143~cm. For both sets of measurements, the ratio $V_m/V_L$ were very stable and the measurement uncertainties were less than the point size used in the plots.
Least-squares fits to Eq.~(\ref{eq:4}) were performed to extract best-fit values of $C_p$ and $C_0^{\prime \prime}$ as a function of $d$ and $h$.  The fit is excellent indicating that the conceptual model for the sensor is accurate.  

{In Figs.~\ref{fig:Fig7}(c) and (d), the measurement results for $C_0$ and $C_p$ with the sensor suspended from the neutral conductor are compared with simulation results.}  The experimental values for $C_0$ were estimated by subtracting \mbox{$C_\mathrm{op}+C_s\approx 6.0$~pF} from the $C_0^{\prime \prime}$ measurements.  {These figures also include simulation results (open triangles) for the case where the sensor is attached to the live conductor.}

\begin{figure}[t] 
\centering
\includegraphics[width=8 cm]{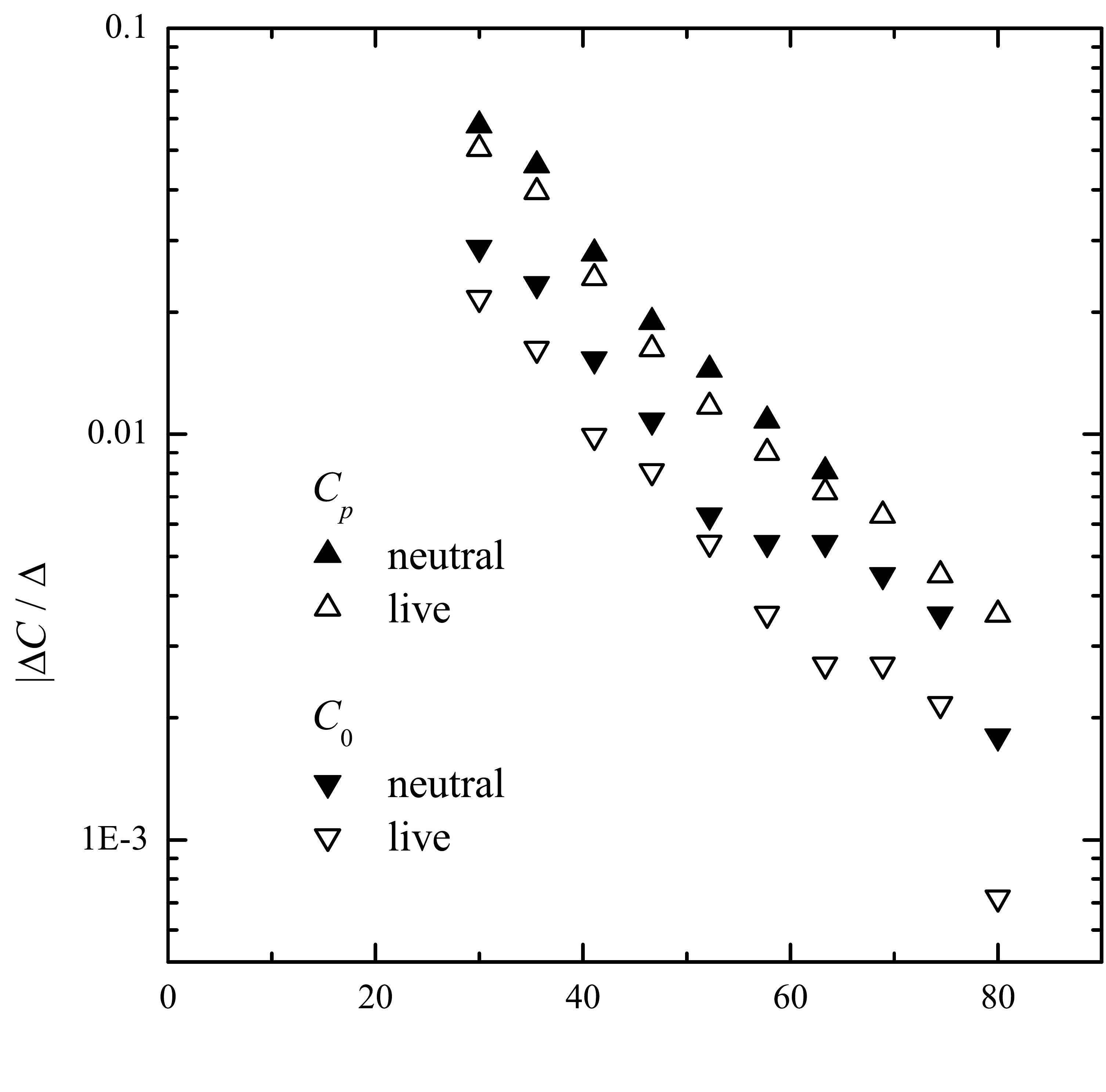} 
\caption{{Plot of the absolute value of the slopes of the simulated $C_p$ and $C_0$ values versus $d$ as a function of the spacing $d$ between the conductors on a semi-log plot.  Filled (open) triangles are for the sensor suspended from the neutral (live) conductor.}}
\label{fig:Fig8}
\end{figure}

There are several important  observations that can be made from these results.  First, there is very good agreement between experimental and numerical results  both in terms of the magnitude of $C_p$ and the dependence of $C_p$ on conductor geometry.  This agreement provides confidence that the simple COMSOL model used in Fig.~\ref{fig:Fig6} produces reliable results.  Second, the simulated results show again that when the sensor is moved from the neutral to the live conductor, $C_p$ is increased and $C_0$ is decreased resulting in an enhancement in the measurement sensitivity.  A third observation is that, in general, $C_0$ is less sensitive to changes in the conductor geometry compared to $C_p$.  This point is clarified in Fig.~\ref{fig:Fig8} where the relative changes to the $C_p$ and $C_0$ values are shown with respect to the conductor spacing $d$.    
The data in Fig.~\ref{fig:Fig8} were obtained by differentiating the capacitance plots in Fig.~\ref{fig:Fig7}(c) and then plotting the magnitude of the derivative (sensitivity) as a function of conductor spacing. The results clearly show that $C_0$ is less sensitive to changes in conductor spacing than $C_p$ regardless of whether the voltage sensor is suspended from the neutral or the live conductor.  {Furthermore, the sensitivity of the net capacitance \mbox{$C_0^{\prime \prime} = C_0 + C_\mathrm{op} + C_s$} is decreased further by the additional contributions of fixed capacitances $C_\mathrm{op}$ and $C_s$ which are independent of $d$.}  Notice also that the sensitivity of both $C_0$ and $C_p$ to changes in  $d$ becomes small for $d\gg L$ which is likely to be the case in most practical applications. (Recall that $L$ is the overall length that the capacitive sensor extends out from the sensor body that houses the electronics.  For the prototype sensor, $L=12.5$~cm.)  Additionally, in any practical application, the conductor spacing would change by only a tiny fraction of its equilibrium spacing and the corresponding change in $C_0^{\prime \prime}$ is likely to be negligible.   We note that both $C_p$ and $C_0$ are even less sensitive to changes in the conductor height.   

Making $C_0$ insensitive to conductor geometry is an important consideration because, as will be emphasized below, the voltage sensor is only reliable if $C_0^{\prime \prime}$ is known accurately via a factory calibration prior to installation.  After installation, small changes to conductor geometry must not result in significant changes to $C_0$.  In a high-voltage transmission line application, seasonal changes in the average temperature will result in changes to the overall conductor length.  Warm summer temperatures will result in the conductors sagging while cool winter temperatures will pull them taut.  If the capacitive sensor is high off of the ground, far away from the non-contacted conductor, and has non-negligible \mbox{$C_\mathrm{op}+C_s$} contributions to $C_0^{\prime \prime}$, then the seasonal changes in the transmission line geometry are expected to result in negligible changes to the overall value of $C_0^{\prime \prime}$.  {We also note that a commercial sensor would be hermetically sealed to desensitise the high-impedance circuitry to changes in humidity.}

\begin{figure}[t] 
\centering
\includegraphics[width=8 cm]{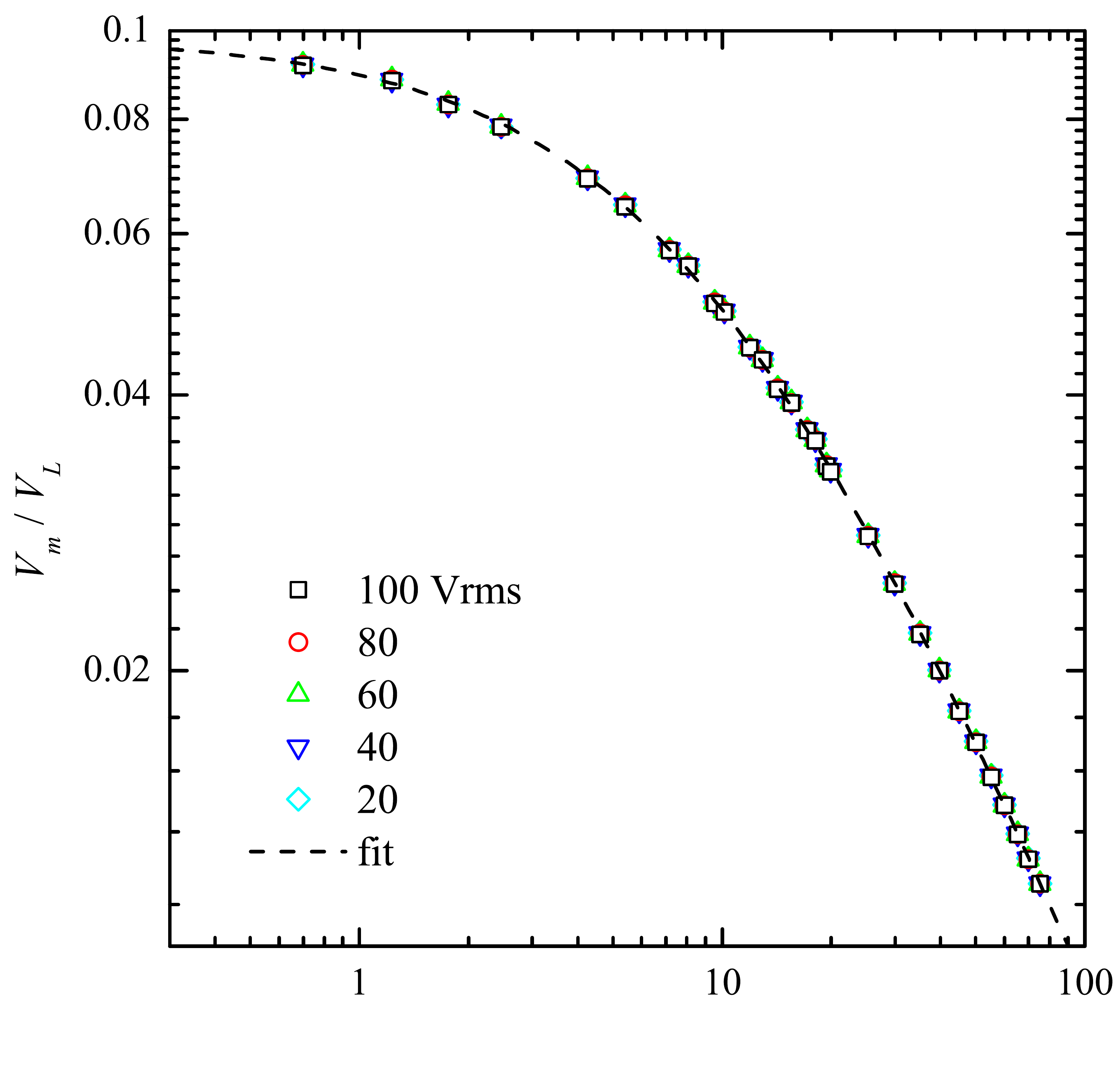} 
\caption{{Plot of $V_m/V_L$ as a function of $C_b$ for five different values of $V_L$ and a least-squares best fit.  The data were obtained using $d=40$~cm and $h=143$~cm.}}
\label{fig:Fig9}
\end{figure}

{For five different values of $V_L$ equally spaced between 20 and 100~Vrms, Fig.~\ref{fig:Fig9} shows $V_m/V_L$ as a function of $C_b$ with $d=40$~cm, $h=143$~cm, and the sensor suspended from the neutral conductor.}  These five datasets are indistinguishable within experimental error and have been simultaneously fitted to Eq.~(\ref{eq:4}) resulting in $C_p=1.00$~pF and $C_0^{\prime \prime}=9.18$~pF.
These data clearly demonstrate that $C_p$ and $C_0^{\prime \prime}$ are independent of the line voltage which is critical for a practical implementation of the voltage sensor.  

\begin{figure*}
\centering
\includegraphics[width=13 cm]{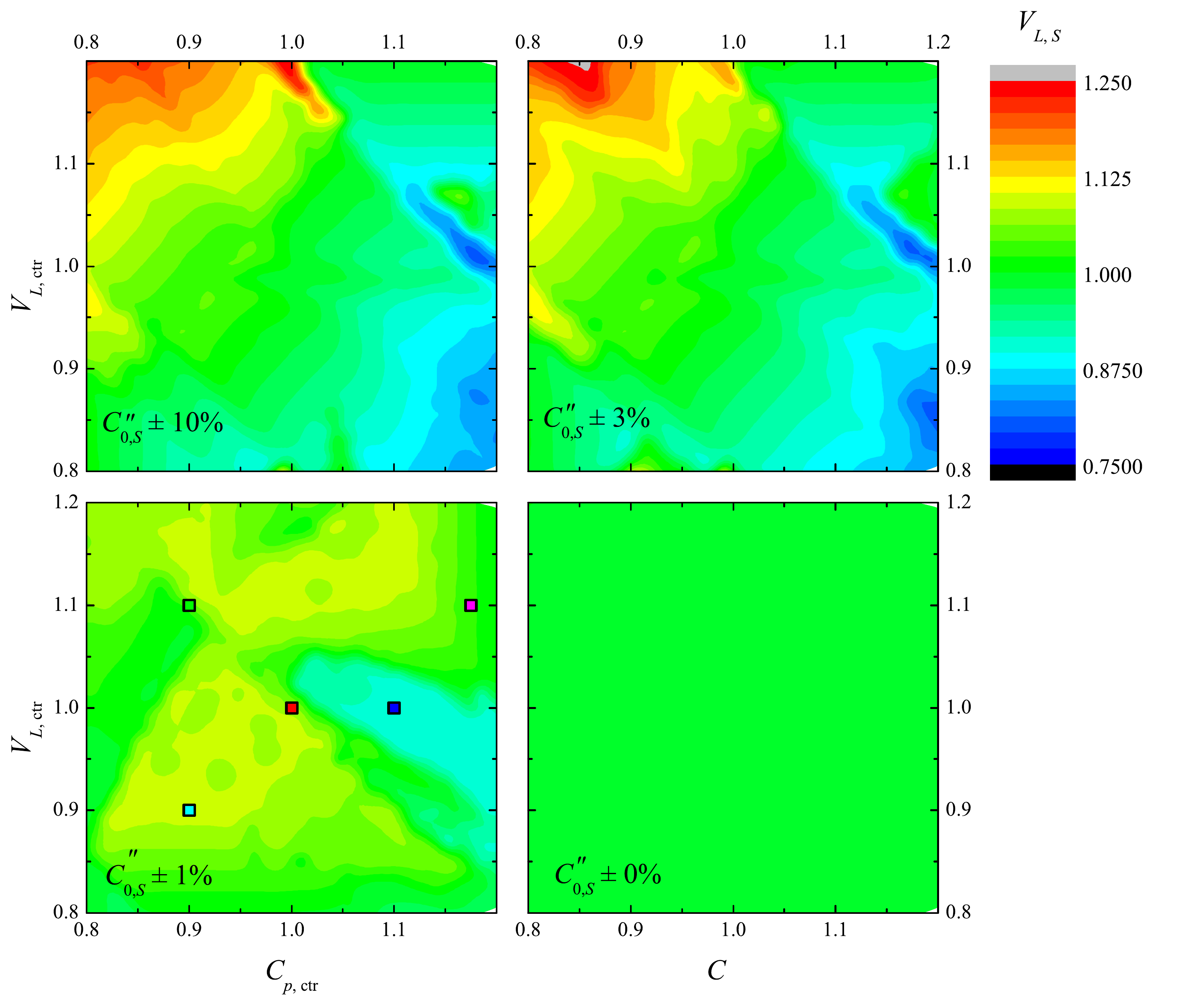} 
\caption{{Analysis of the extracted line voltage $V_L$ as a function of the parameter ranges used for $V_L$ and $C_p$ in the nonlinear least-squares fitting routine.  The analysis was repeated allowing the fit range of the $C^{\prime \prime}_0$ parameter to be $\pm 10$\%, 3\%, 1\%, and 0\%. The squares in the $\pm1$\% plot indicate values of $C_{p,\mathrm{ctr}}$ and $V_{L,\mathrm{ctr}}$ for which the extracted line voltage was plotted versus the known line voltage in Fig.~\ref{fig:Fig11}.}}
\label{fig:Fig10}
\end{figure*}
Thus far, all of the $V_m$ data presented from the prototype sensor have been obtained with an unterminated (open) transmission line and therefore a line current of zero.  To demonstrate that $V_m$, and therefore $C_p$ and $C_0^{\prime \prime}$, are also independent of the line current; the $V_m/V_L$ versus $C_b$ measurements using $V_L=100$~Vrms were repeated with the transmission line terminated with a 60~W incandescent light bulb, a $6~\mu$F capacitor, and an ac electric motor.  These three terminations caused the magnitude of the line current to range between 0.24 and 0.59~A and the phase to vary from approximately $-\pi/2$ to $\pi/2$~radians relative to the line voltage.  Within experimental error, there was no change in the measured $V_m/V_L$ values for all three of these terminations relative to an open transmission line.

{Finally, in order to measure the bandwidth of the sensor, the testbed transmission line was driven by a function generator and the outputs of the generator and the voltage sensor were simultaneously displayed on an oscilloscope.  With \mbox{$C_0^{\prime \prime}+C_p\approx 10$~pF}, the low-frequency corner frequency is expected to be less than 1~mHz.  The sensor output $V_m$ was measured as a function of frequency from 0.1~Hz to 100~kHz and found to be frequency independent and in phase with the generator voltage from 0.1~Hz to 20~kHz.  Above 20~kHz, $V_m$ increasingly lags behind the generator voltage as the gain-bandwidth product of the op-amp is approached.  Bandwidth measurements were preformed using bank capacitances of $0$ and $100$~pF and the two sets of measurements were indistinguishable within experimental error.}

\subsection{Blind Estimation of Line Voltage}\label{sec:abs}

In Fig.~\ref{fig:Fig9}, the measured $V_m$ were scaled by the known line voltages $V_L$ and then fitted to Eq.~(\ref{eq:4}) to extract reliable values for $C_p$ and $C_0^{\prime \prime}$.  In this section, a blind estimation of the line voltage $V_L$ is made from measurements of $V_m$ as a function of $C_b$.  As previously discussed in section~\ref{sec:analysis}, either $C_p$ or $C_0^{\prime \prime}$ must be known in order to fit the data to a unique set of parameters.  Because $C_0^{\prime \prime}$ is relatively insensitive to small changes in conductor geometry, it is feasible to determine $C_0^{\prime \prime}$ accurately from a factory calibration.  In this work, the calibration procedure uses fits to the data presented in Fig.~\ref{fig:Fig9} to directly obtain the value of $C_0^{\prime \prime}$.  

To demonstrate how the voltage sensor can be used to extract the line voltage, the  data in Fig.~\ref{fig:Fig9} measured for a line voltage of $V_L=100$~Vrms were fitted to a scaled version of Eq.~(\ref{eq:4}):
\begin{equation}
V_m=\frac{\left(V_L V_{L,S}\right)\left(C_p C_{p,S}\right)}{C_0^{\prime \prime}C_{0,S}^{\prime \prime}+C_b+C_p C_{p,S}}
\label{eq:6}
\end{equation}
where \mbox{$C_p=1.00$~pF} and \mbox{$C_0^{\prime \prime}=9.18$~pF} are the known values from the fit in Fig.~\ref{fig:Fig9} and $V_{L,S}$, $C_{p,S}$, and $C^{\prime \prime}_{0,S}$ are scaled fit parameters.  In the nonlinear best-fit routine, the allowable range of each fit parameter must be specified.  For example, the allowable range for the scaled line voltage would be specified as \mbox{$V_{L,\mathrm{ctr}}-\Delta V_{L,S}\le V_{L,S}\le V_{L,\mathrm{ctr}}+\Delta V_{L,S}$} where $V_{L,\mathrm{ctr}}$ is the center of the range and $2\Delta V_{L,S}$ is the span of the range.  The 100~Vrms data of Fig.~\ref{fig:Fig9} were {fitted} to Eq.~(\ref{eq:6}) using \mbox{$0.8\le V_{L,\mathrm{ctr}},C_{p,\mathrm{ctr}}\le 1.2$} in steps of 0.005 and with \mbox{$\Delta V_{L,S}=\Delta C_{p,S}=0.20$} resulting in $6561$ unique parameter ranges and corresponding best-fit parameters. For all fits, $C^{\prime \prime}_{0,\mathrm{ctr}}=1$ was used.  The set of $6561$ fits was repeated using $\Delta C_{0,S}^{\prime \prime}=0.10$, $0.03$, $0.01$, and zero.  False-color plots of the extracted $V_{L,S}$ values as a function of $V_{L,\mathrm{ctr}}$ and $C_{p,\mathrm{ctr}}$ are presented in Fig.~\ref{fig:Fig10}. 

These results show that the extracted line voltages are typically within 10\% of the known value, falling outside of this range only when the $C_0^{\prime \prime}$ fit range is relatively large and either $V_{L,\mathrm{ctr}}$ or $C_{p,\mathrm{ctr}}$ is significantly less than one while the other is simultaneously significantly greater than one.  As discussed in section~\ref{sec:analysis}, any set of parameters \mbox{$\{V_{L,S},C_{p,S},C^{\prime \prime}_{0,S}\}$} that preserves the product \mbox{$V_{L,S} C_{p,S}$} and the sum \mbox{$C_{0,S}^{\prime \prime}+C_{p,S}$} will produce identical fits to the data.  Consider, for example, the case $V_{L,\mathrm{ctr}}>1$ and $C_{p,\mathrm{ctr}}<1$, but with \mbox{$V_{L,\mathrm{ctr}}C_{p,\mathrm{ctr}}\approx 1$}.  A good fit to the data can be achieved with these initial parameters provided that $C_{0,S}^{\prime \prime}$ has sufficient range to compensate for the reduced value of $C_{p,\mathrm{ctr}}$.  This case is shown in the top left-hand corners of Fig.~\ref{fig:Fig10} when $\Delta C_{0,S}^{\prime \prime}=0.10$ and $0.03$ resulting in $V_{L,S}>1$ (red/orange coloring).  However, if $C_{0,S}^{\prime \prime}$ is known very accurately such that $\Delta C_{0,S}^{\prime \prime}$ is restricted, then $C_{0,S}^{\prime \prime}$ is unable to compensate for the reduced value of $C_{p,\mathrm{ctr}}$ and the parameters $V_{L,\mathrm{ctr}}$ and $C_{p,\mathrm{ctr}}$ must be adjusted closer to their true values in order to achieve a good fit to the data.  This case is shown in Fig.~\ref{fig:Fig10} when $\Delta C_{0,S}^{\prime \prime}=0.01$ and zero.

We now quantify the bound on $\Delta C_{0}^{\prime \prime}$ required to guarantee that the $V_L$ extracted from the fitting procedure is within $\Delta V_L$ of the true value.  The fitting constraint $V_{L,S}C_{p,S}=1$ implies that $\Delta C_{p,S}\approx \Delta V_{L,S}$ since $V_{L,S}\approx 1$. The second requirement for a good fit is:
\begin{equation}
C_0^{\prime \prime}+C_p=C_{0}^{\prime \prime}C_{0,S}^{\prime \prime}+C_pC_{p,S}\label{eq:7}
\end{equation}
which can be solved for $C_{0,S}^{\prime \prime}$.  Applying propagation of errors results in:
\begin{equation}
\Delta C_{0,S}^{\prime \prime}\approx\frac{C_p}{C_0^{\prime \prime}}\Delta V_{L,S}\label{eq:8}
\end{equation}
which is consistent with the sensor sensitivity being approximately proportional to $C_p/C_0^{\prime \prime}$.  

\begin{figure}[t] 
\centering
\includegraphics[width=8 cm]{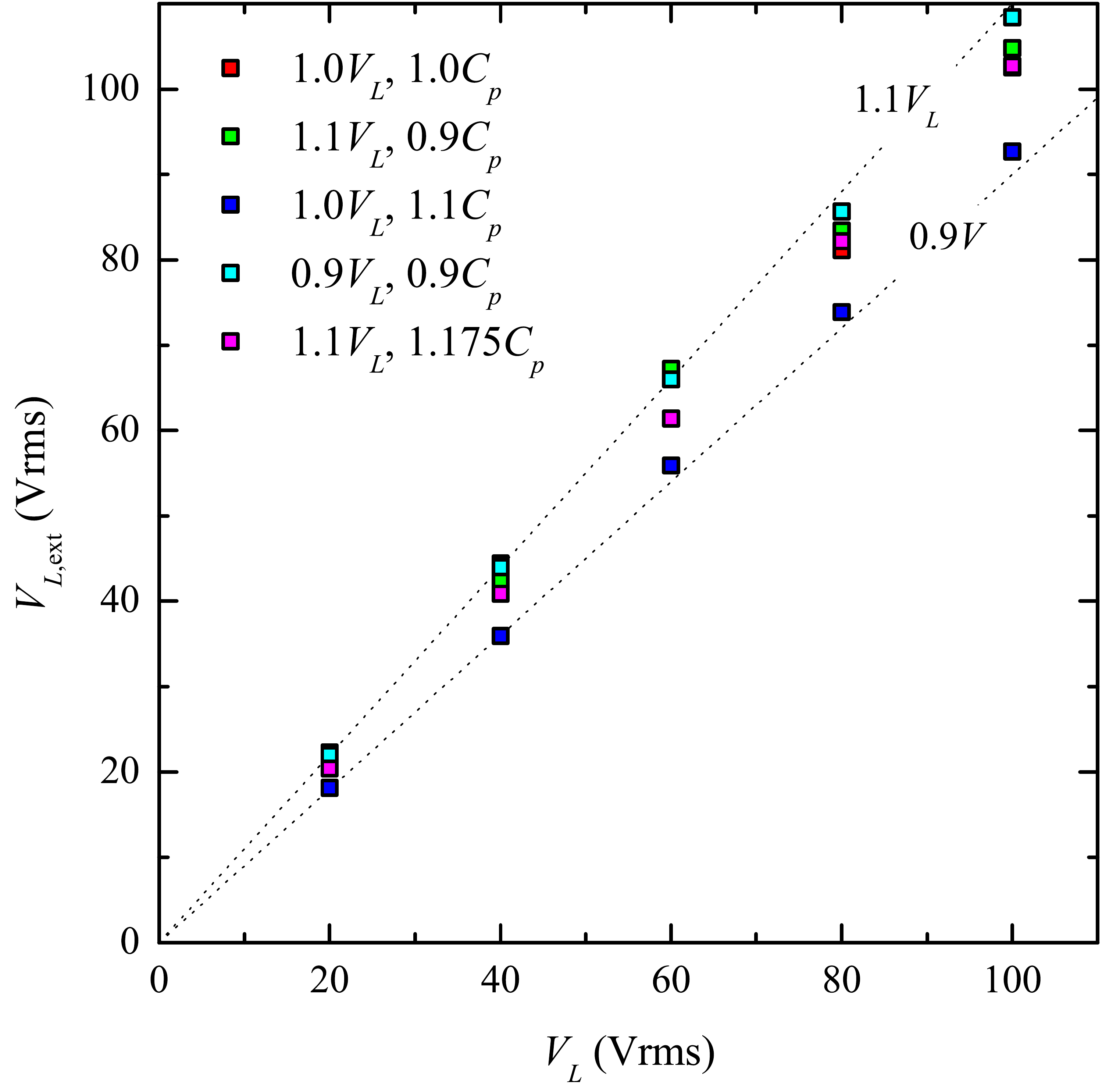}
\caption{{The line voltage extracted from the nonlinear regression analysis plotted as a function of the known line voltage.  The colored squares correspond to those shown in Fig.~\ref{fig:Fig10}.}}
\label{fig:Fig11}
\end{figure}

For the prototype sensor, $C_p=1.00$~pF and $C_0^{\prime \prime}=9.18$~pF, such that $C_0^{\prime \prime}$ must be known within 2.2\% of its true value to determine $V_L$ to within 20\% of its known value.  Two of the plots in Fig.~\ref{fig:Fig10} were generated using $\Delta C_{0,S}^{\prime \prime}>2\%$ and result in a $V_{L,S}$ with an accuracy worse than $\pm 20$\%.  On the other hand, the two plots for which $\Delta C_{0,S}^{\prime \prime}<2\%$ result in $V_{L,S}$ and accuracy which is better than $\pm 20$\%.  In particular, with $\Delta C_{0,S}^{\prime \prime}=1\%$, the accuracy of the line voltage estimate is approximately 10\% which is consistent with Eq.~(\ref{eq:8}).  This point is emphasized in Fig.~\ref{fig:Fig11} which shows the extracted line voltage $V_{L,\mathrm{ext}}$ from blind analyses of $V_m$ versus $C_b$ datasets as a function of the known line voltage.  The data in Fig.~\ref{fig:Fig11} were generated for applied line voltages from 20 to 100~Vrms and assuming that $C_0^{\prime\prime}$ was known to with 1\% of its true value. The different colored squares in Fig.~\ref{fig:Fig11} indicate the center of the parameter ranges used for $V_L$ and $C_p$ in the nonlinear fits and correspond to colored squares shown in Fig.~\ref{fig:Fig10}.  The data in Fig.~\ref{fig:Fig11} clearly show that the extracted line voltages always fall within 10\% of the actual line voltages.

{We emphasize that the 10\% accuracy shown in Fig.~\ref{fig:Fig11} is merely a demonstration of using to the prototype sensor to extract line voltages assuming that a factory calibration has determined $C_0^{\prime \prime}$ to within 1\% of its true value.  The accuracy of the estimated $V_L$ could be improved in three ways.  First, the analyses presented in Figs.~\ref{fig:Fig10} and \ref{fig:Fig11} were done with the voltage sensor suspended from the neutral conductor.  By simply moving the sensor to the high-voltage conductor, Table~\ref{tab:Tab1} shows that $C_p/C_0^{\prime \prime}$ is expected to increase by nearly a factor of two which would result in a decrease in $\Delta V_{L,S}$ by the same factor.  Second, the  $C_p/C_0^{\prime \prime}$ ratio can be further increased by optimizing the circuit layout to the enhance the suppression of shunt capacitance ($C_s$) contributions to $C_0^{\prime \prime}$.  Finally, the most effective way to improve the sensor accuracy is by using precise factory calibrations to reduce the relative uncertainty in $C_{0}^{\prime \prime}$ below 1\%.  A precision calibration would likely involve determining the temperature dependence of $C_0^{\prime \prime}$ and then equipping the voltage sensor with an on-board temperature sensor.  This capability would allow for real-time corrections to the calibration values of $C_0^{\prime \prime}$ due to daily and seasonal temperature fluctuations.}

\section{High-Voltage Applications}\label{sec:HV}

Up to this point, all of the data that has been presented has been limited to line voltages of 100~Vrms or less.  Since the main motivation for designing the sensor is to measure high ac voltages, the design needs to be scaled and verified under high-voltage conditions.

The prototype sensor described so far uses a capacitive divider between the two conductors and the measured  voltage at the output of the buffer is given by \mbox{$V_m=V_LC_p\left(C_0^{\prime \prime}+C_p\right)^{-1}\approx V_L/10$} when $C_b=0$.  Because the buffer is powered by a low-voltage supply ($\pm 18$~V), its output saturates for line voltages greater than 120~Vrms. The voltage division across the capacitor divider is determined by $C_p$ and $C_0^{\prime \prime}$ and, for reasons given earlier, it is important to have a low ratio to maintain high accuracy. On the other hand, the low ratio means the voltage division {factor} is small and consequently a high voltage is measured at the input of the op-amp. 

\begin{figure}[t] 
\centering
\includegraphics[width=\columnwidth]{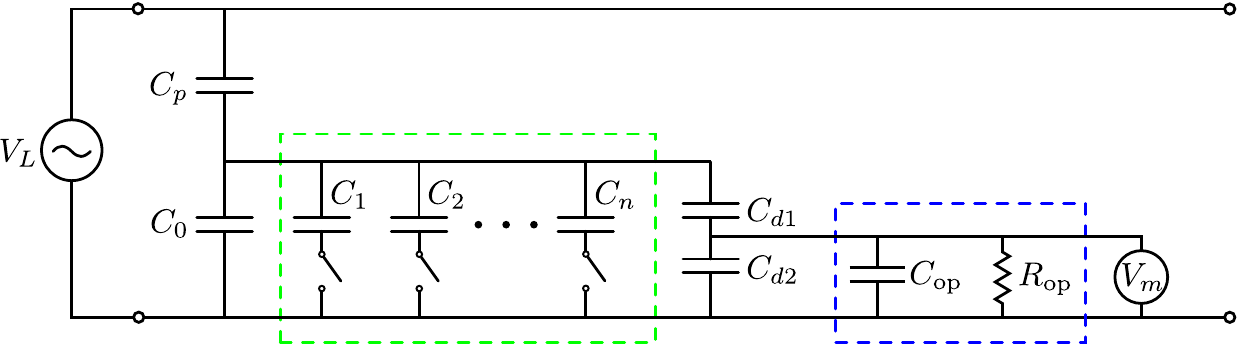}
\caption{The voltage sensor configured for high-voltage applications.  An additional capacitive divider is used between the capacitor bank and the buffer input.}
\label{fig:Fig12}
\end{figure}

As a way to maintain accuracy and increase the voltage division ratio, Fig.~\ref{fig:Fig12} shows a design modification that enables the sensor to be scaled to high-voltage applications.  A two-step voltage divider is set-up using two additional capacitors, $C_{d1}$ and $C_{d2}$.  Neglecting the input resistance of the op-amp, the measured output is given by:     
\begin{equation}
\begin{split}
V_m&=V_L\left[\frac{C_p}{C_0+C_s+\left(\dfrac{1}{C_{d1}}+\dfrac{1}{C_{d2}+C_\mathrm{op}}\right)^{-1}+C_b+C_p}\right]\\
&\qquad\times\left(\frac{C_{d1}}{C_{d1}+C_{d2}+C_\mathrm{op}}\right)
\end{split}
\label{eq:9}
\end{equation}   
where $C_s$ is the stray capacitance associated with the layout of the circuit and is not shown in Fig.~\ref{fig:Fig12}.  In high-voltage applications, the condition $C_{d2}\gg C_{d1}$ is required to significantly step the voltage down, therefore:
\begin{equation}
V_m\approx V_L\left(\frac{C_p}{C_{0,\mathrm{HV}}^{\prime \prime}+C_b+C_p}\right)\frac{C_{d1}}{C_{d2}}
\label{eq:10}
\end{equation}
where \mbox{$C_{0,\mathrm{HV}}^{\prime \prime}=C_0+C_s+C_{d1}$}.
If $C_{d1}$ is selected to be of the same order as $C_\mathrm{op}\approx 0.8$~pF, then \mbox{$C_{0,\mathrm{HV}}^{\prime \prime}\approx C_0^{\prime \prime}$}. {Under this condition, $V_m/V_L$ is reduced by an additional factor of $C_{d2}/C_{d1}$ without significantly altering either the accuracy or bandwidth of the sensor.} The ratio $C_{d2}/C_{d1}$ can be made arbitrarily large and the primary limitation becomes the dielectric breakdown voltage of the capacitors.  

\begin{figure}[t] 
\centering
\includegraphics[width=8 cm]{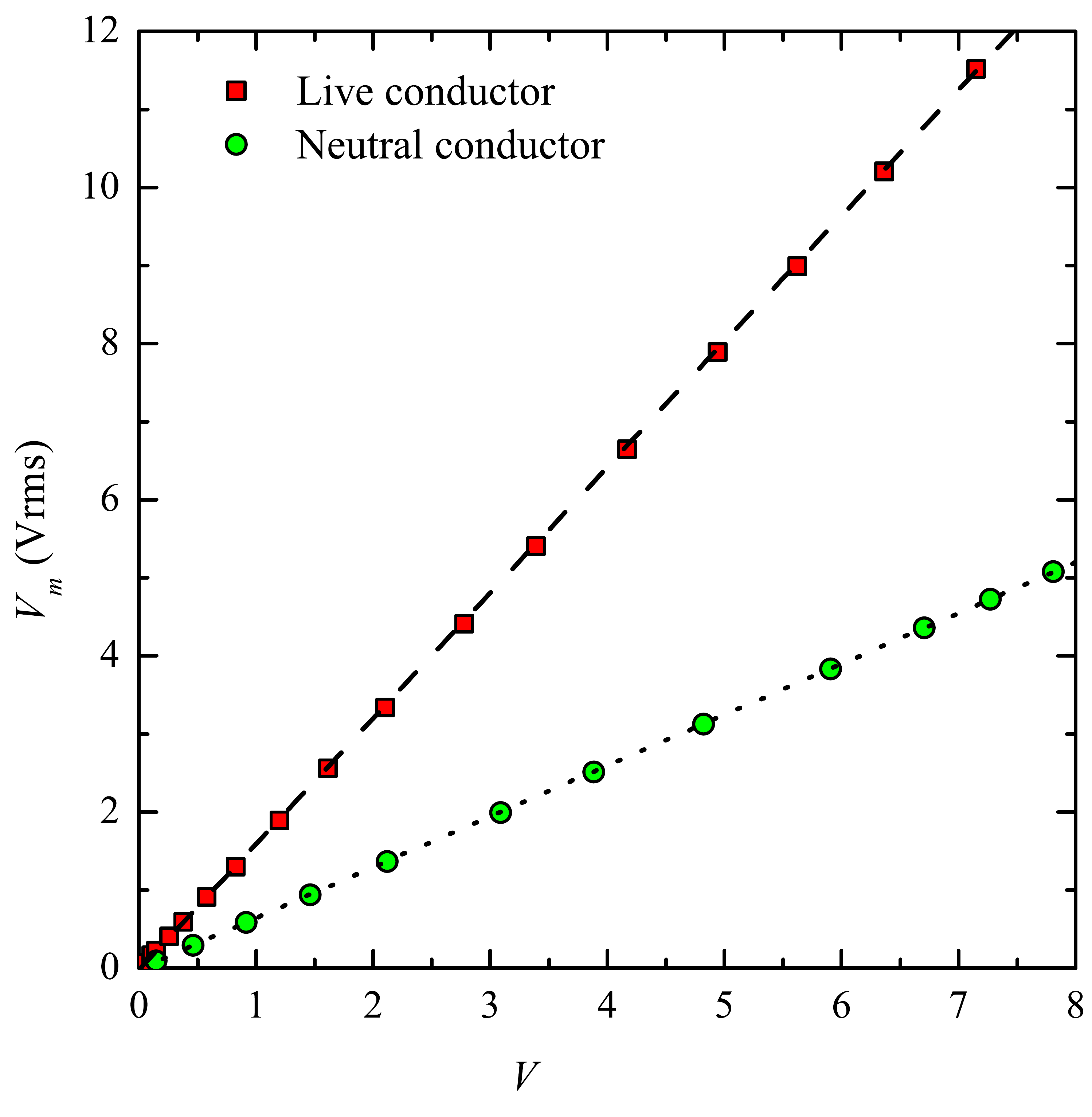} 
\caption{{$V_m$ plotted as a function of $V_L$ with the voltage sensor suspended from the neutral conductor (circular points) and live conductor (square points) while configured for high-voltage measurements.}}
\label{fig:Fig13}
\end{figure}

The two-stage voltage divider shown in Fig.~\ref{fig:Fig12} was implemented in the prototype sensor to verify the circuit. In the circuit{,} nominal capacitances of $C_{d1}=0.5$~pF and $C_{d2}=130$~pF were used.  A step-up transformer with a turns ratio of \mbox{62.5:1} was used with a 0--140~Vrms variac feeding the primary winding. With this configuration, the line voltage could be increased up to 7.5~kVrms. A high-voltage probe (Tektronix P6015A) was used to measure the exact line voltage.  The picture shown earlier in  Fig.~\ref{fig:Fig5} includes the two-stage voltage divider and {the 62.5-times step-up} transformer. {Figure~\ref{fig:Fig13} shows the measured $V_m$ data plotted as a function of $V_L$ up to 7.5~kVrms with the voltage sensor suspended from the neutral conductor and suspended from the live conductor.}
{As expected, both data sets are very linear ($R^2=1.00$).  These datasets also experimentally confirm that suspending the voltage sensor from the live conductor results in an enhanced measurement sensitivity (increased slope).}

\section{Conclusion}\label{sec:conclusions}

A sensor capable of accurately measuring the ac voltage across a pair of conductors with a single contact to one conductor has been described and evaluated experimentally. The sensor uses the stray capacitance $C_p$ from a sensing plate to the non-contacted conductor to create a capacitive voltage divider that can be used to measure the line voltage. A straightforward factory calibration to determine the net capacitance $C_0^{\prime \prime}$ from the sensing plate back to the main body of the sensor is required to extract the absolute line voltage.  A set of measurements with calibrated bank capacitors are then made and a nonlinear constrained optimizer is used to find a set of best-fit parameters, which includes an estimate of the unknown line voltage.  {With the prototype sensor suspended from the neutral conductor, if $C_0^{\prime \prime}$ is known to within 1\% of its true value, then the absolute line voltage was determined to within 10\% of its true value. Moving the sensor to the live conductor, optimizing the circuit layout to minimize shunt capacitances to the op-amp input, and refined factory calibrations of $C_0^{\prime \prime}$ are all practical methods that can be used to improve the accuracy of the estimated line voltage.}  

Other important contributions of this work include an investigation into the best shape of the capacitor sensing plate. Four different designs were evaluated using numerical simulations and a flat disk sensor was found to be the best design. Simulation results were used to explore the difference between mounting the sensor on the neutral or live (high-voltage) line. {The simulation results, supporting theory, and experimental measurements all conclude that better sensitivity and accuracy can be obtained by contacting the high-voltage conductor.} Finally, a two-stage capacitor divider was tested to increase the operating voltage of the sensor up to 7.5~kVrms. Future work will include testing the design up to much higher line voltages and investigating three-phase applications. 

\section*{Acknowledgment}

The authors gratefully acknowledge the support of Awesense Wireless, NSERC, and CMC Microsystems.

\ifCLASSOPTIONcaptionsoff
  \newpage
\fi



\bibliographystyle{IEEEtran}
\bibliography{voltage_sensor_refs_v1}
%



%

\begin{IEEEbiography}[{\includegraphics[width=1in,height=1.25in,clip,keepaspectratio]{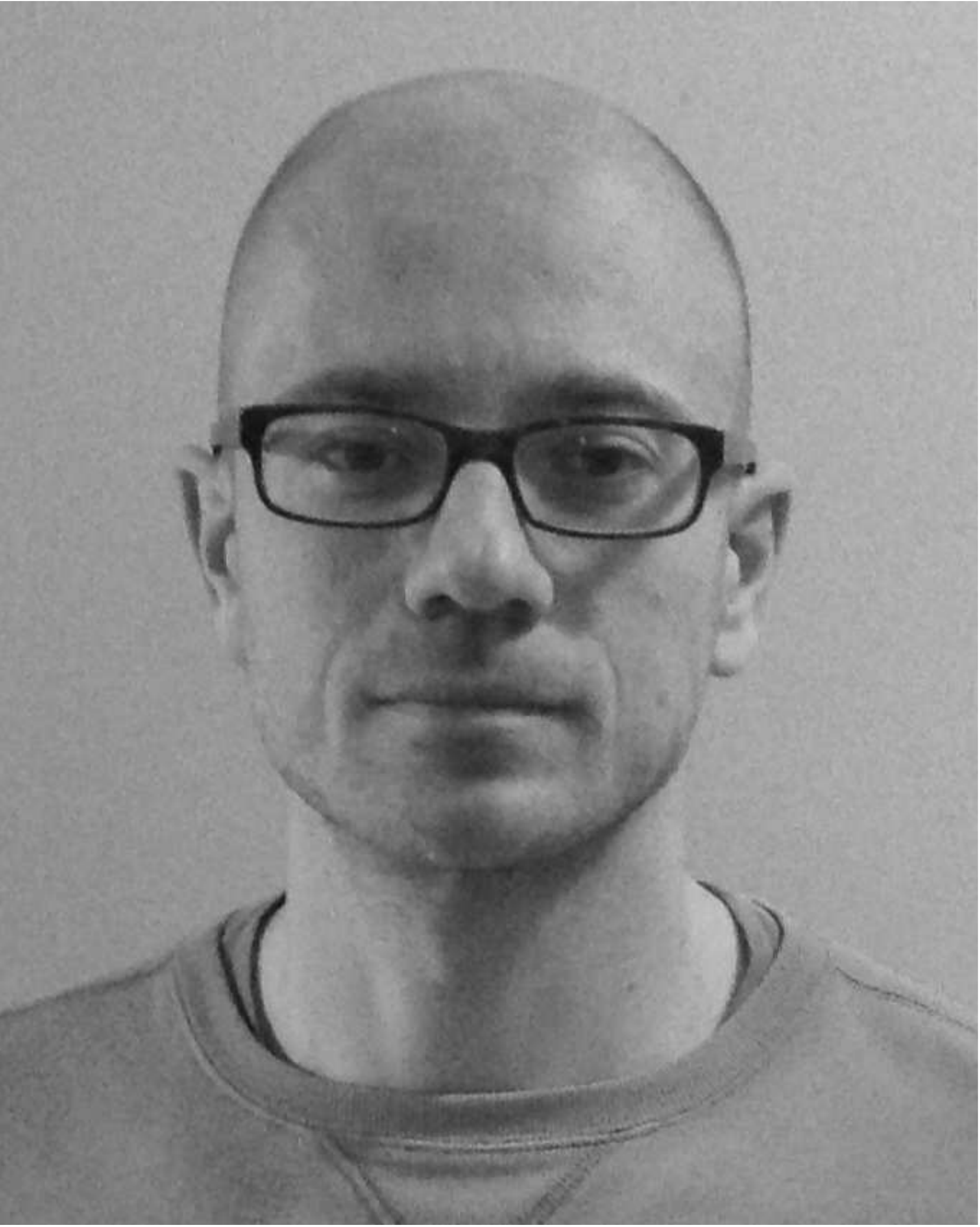}}]{Jake S. Bobowski}
  was born in Winnipeg, Canada on March 9, 1979.  He received a B.Sc. degree in physics from the University of Manitoba, Canada in 2001.  He was awarded M.Sc. and Ph.D. degrees in physics from the University of British Columbia, Canada in 2004 and 2010, respectively.  From 2011 to 2012, he was a postdoctoral fellow in the Department of Electrical Engineering, and is now a physics instructor, at the Okanagan campus of the University of British Columbia, Canada. 
\end{IEEEbiography}

\begin{IEEEbiography}[{\includegraphics[width=1in,height=1.25in,clip,keepaspectratio]{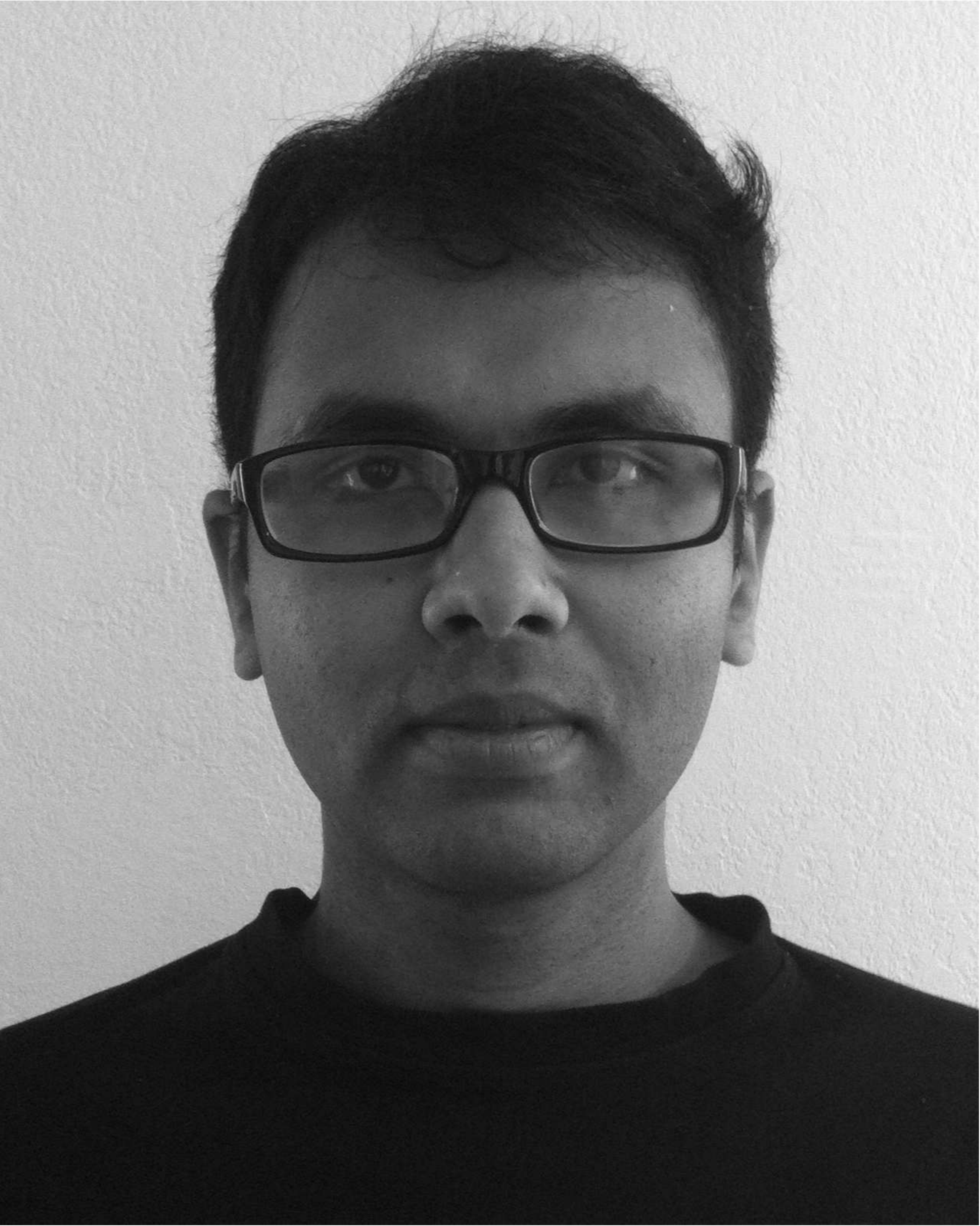}}]{Saimoom~Ferdous}
  was born in July, 1988 in Rangpur, Bangladesh.  He received his B.Sc. degree in electrical and electronic engineering from the Bangladesh University of Engineering and Technology, Dhaka, Bangladesh in 2012.  Currently, he is pursuing a M.A.Sc. degree in electrical engineering at the University of British Columbia, Okanagan campus, Canada.  His research interests include the development of new electromagnetic devices for biomedical, agricultural, and remote sensing applications.
\end{IEEEbiography}

\begin{IEEEbiography}[{\includegraphics[width=1in,height=1.25in,clip,keepaspectratio]{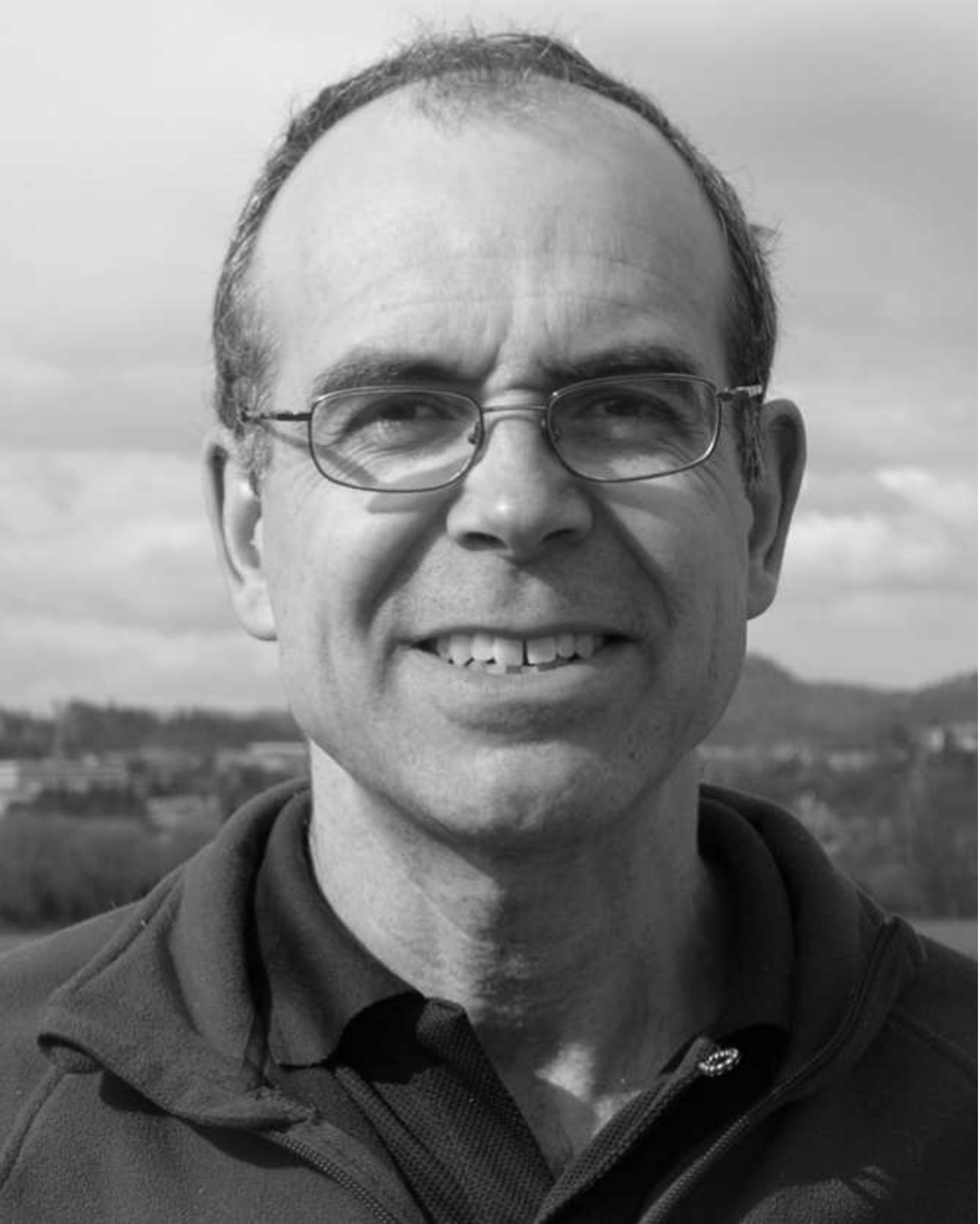}}]{Thomas~Johnson}
is an Assistant Professor in the School of Engineering at the University of British Columbia. His research interests include the design and implementation of radio frequency circuits and systems and applications of electromagnetic field theory. Before joining UBC in 2009, he worked as a technical lead in a number of companies including PulseWave RF, ADC Telecommunications, and Norsat International. He has Ph.D. (2007) and MASc (2001) degrees from Simon Fraser University.
\end{IEEEbiography}





\end{document}